\documentclass[11pt,a4paper]{article}

\usepackage[utf8]{inputenc}
\usepackage[left=25mm,right=25mm,top=20mm,bottom=20mm]{geometry}

\usepackage{xspace}
\usepackage{etoolbox}

\usepackage{tocbibind}
\usepackage{subfiles}
\usepackage{subfiles-commands}

\newcommand\subfileref{%
  \onlyinsubfile{%
    \printbibliography
  }%
}

\usepackage{relsize}
\usepackage{graphicx}
\usepackage[svgnames]{xcolor}
\usepackage{placeins}
\usepackage{float}
\usepackage{subfigure}
\usepackage{caption}

\usepackage{tikz}
\usepackage{tikz-relcoord}
\usetikzlibrary{arrows.meta,positioning,backgrounds}
\usetikzlibrary{matrix}

\usepackage{framed} 
\colorlet{shadecolor}{yellow!50!white}

\usepackage{amsmath,amssymb}
\usepackage[uncertainty-mode=separate]{siunitx}
\usepackage{wasysym}
\usepackage{tensor}

\usepackage{array}
\usepackage{booktabs}
\usepackage{multirow}
\usepackage{tabulary}
\usepackage[affil-it]{authblk}

\usepackage[color={1.0 0.9 0.5}]{pdfcomment}
\hypersetup{colorlinks,
linkcolor=blue,
anchorcolor=blue,
citecolor=blue,
urlcolor=blue}

\usepackage{lineno}
\usepackage{lineno-ams}
\usepackage[normalem]{ulem}
\usepackage{url}
\usepackage[numbers,sort&compress]{natbib}
\usepackage[capitalise,noabbrev]{cleveref}

\usepackage{docdbref}
\usepackage{forcsvlistand}

\newcommand\printbibliography{
  \bibliographystyle{h-physrev5}
  \bibliography{references}
}

\newcommand*\Dm[2]{\ensuremath{\Delta m^2_{#1#2}}}

\newcommand*\SinSqTheta[2]{\ensuremath{\sin^2\theta_{#1#2}}}
\newcommand*\SinSqDTheta[2]{\ensuremath{\sin^22\theta_{#1#2}}}

\xspaceaddexceptions{;}

\newcommand\Min{\ensuremath{\operatorname{min}\xspace}}
\newcommand\DTestStat{\ensuremath{\Delta{}{\cal T}}\xspace}
\newcommand\DTestStatA{\ensuremath{\overline{\Delta{}{\cal T}}}\xspace}
\newcommand\TestStat[1][]{\ensuremath{{\cal T}_{\mathrm{#1}}}\xspace}
\newcommand\TestStatC{\ensuremath{{\cal T}_C}\xspace}

\newcommand{\SensN}{\ensuremath{n}\xspace}

\newcommand\Data[1][]{\ensuremath{D_\mathrm{#1}}\xspace}
\newcommand\DataA[1][]{\ensuremath{\overline{D}_\mathrm{#1}}\xspace}
\newcommand\Datai{\ensuremath{D}\xspace}
\newcommand\Theory{\ensuremath{\mu}\xspace}

\newcommand*\DChi{\ensuremath{\Delta \chi^2}\xspace}

\newcommand*\ChiSqOsc{\ensuremath{\chi^2_\text{osc}}\xspace}
\newcommand*\ChiSqNuisance{\ensuremath{\chi^2_\text{nuis.}}\xspace}
\newcommand*\ChiSqNuisanceCorr{\ensuremath{\chi^2_\text{corr.nuis.}}\xspace}

\newcommand*\tstat{\ensuremath{_\text{stat}}\xspace}
\newcommand*\tbtob{\ensuremath{_\text{b2b}}\xspace}
\newcommand*\IdxDet{d}
\newcommand*\TDet{\ensuremath{^\IdxDet}\xspace}

\newcommand*\ParFree{\ensuremath{\vartheta}\xspace}

\newcommand*\ParNuisance{\ensuremath{\eta}\xspace}
\newcommand*\ParNuisanceCorr{\ensuremath{\zeta}\xspace}

\newcommand*\ParsFree{\ensuremath{\vec{\ParFree}}\xspace}
\newcommand*\ParsNuisance{\ensuremath{\vec{\ParNuisance}}\xspace}
\newcommand*\ParsNuisanceCorr{\ensuremath{\vec{\ParNuisanceCorr}}\xspace}

\newcommand*\Erec{\ensuremath{E^\text{rec}}\xspace}%
\newcommand*\Edep{\ensuremath{E^\text{dep}}\xspace}%
\newcommand*\Evis{\ensuremath{E^\text{vis}}\xspace}%
\newcommand*\Epos{\ensuremath{E^{e}}\xspace}%
\newcommand*\Enu{\ensuremath{E^\nu}\xspace}%
\newcommand*\Ceres{\ensuremath{C^\text{Eres}}\xspace}%

\newcommand\NewText[1]{\texorpdfstring{{\color{blue}#1}}{#1}}

\begin{document}
\title{Potential to identify neutrino mass ordering with reactor antineutrinos at JUNO}
\author[6,5]{Angel Abusleme}
\author[44]{Thomas Adam}
\author[65]{Shakeel Ahmad}
\author[65]{Rizwan Ahmed}
\author[54]{Sebastiano Aiello}
\author[65]{Muhammad Akram}
\author[65]{Abid Aleem}
\author[20]{Fengpeng An}
\author[22]{Qi An}
\author[54]{Giuseppe Andronico}
% ...
\affil[5]{Millennium Institute for SubAtomic Physics at the High-energy Frontier (SAPHIR), ANID, Chile}
\affil[6]{Pontificia Universidad Cat\'{o}lica de Chile, Santiago, Chile}
% ...
\affil[20]{Sun Yat-Sen University, Guangzhou, China}
% ...
\affil[22]{University of Science and Technology of China, Hefei, China}
% ...
\affil[44]{IPHC, Universit\'{e} de Strasbourg, CNRS/IN2P3, F-67037 Strasbourg, France}
% ...
\affil[54]{INFN Catania and Dipartimento di Fisica e Astronomia dell Universit\`{a} di Catania, Catania, Italy}
% ...
\affil[65]{Pakistan Institute of Nuclear Science and Technology, Islamabad, Pakistan}
% ...


\subfilemain

\maketitle

\begin{abstract}
The Jiangmen Underground Neutrino Observatory (JUNO) is a multi-purpose neutrino experiment under construction in South China.
This paper presents an updated estimate of JUNO's sensitivity to neutrino mass ordering using the reactor antineutrinos emitted from eight nuclear reactor cores in the Taishan and Yangjiang nuclear power plants.
This measurement is planned by studying the fine interference pattern caused by quasi-vacuum oscillations in the oscillated antineutrino spectrum at a baseline of 52.5~km and is completely independent of the CP violating phase and neutrino mixing angle $\theta_{23}$.
The sensitivity is obtained through a joint analysis of JUNO and Taishan Antineutrino Observatory (TAO) detectors utilizing the best available knowledge to date about the location and overburden of the JUNO experimental site, local and global nuclear reactors, JUNO and TAO detector responses,
expected event rates and spectra of signals and backgrounds, and systematic uncertainties of analysis inputs.
We find that a 3$\sigma$ median sensitivity to reject the wrong mass ordering hypothesis can be reached with an exposure to approximately 6.5 years $\times$ 26.6~GW thermal power.

\end{abstract}

\tableofcontents


\section{Introduction} 
\label{sec:intro}
The discovery of neutrino oscillations~\cite{Kajita:2016cak,McDonald:2016ixn} is often considered one of the most direct pieces of evidence of new physics beyond the Standard Model of particle physics.
The solar neutrino deficit problem~\cite{Bahcall:1976zz}, first observed by the Homestake experiment~\cite{Davis:1968cp}, was decisively resolved by the SNO experiment in 2002 using simultaneous measurements of charged and neutral current interactions on deuterium, as well as the elastic scattering process on electrons~\cite{SNO:2001kpb,SNO:2002tuh}. This was later confirmed by the KamLAND long-baseline reactor $\bar{\nu}_{e}$ disappearance experiment~\cite{KamLAND:2002uet}. Meanwhile, the atmospheric neutrino anomaly observed by several experiments~\cite{LoSecco:1985py,Kamiokande-II:1988sxn,Becker-Szendy:1992zlj} was unambiguously resolved by the Super-Kamiokande experiment in 1998~\cite{Super-Kamiokande:1998kpq} and later confirmed by the K2K long-baseline accelerator ${\nu}_{\mu}$ disappearance experiment~\cite{K2K:2002icj}. 
Both the solar and atmospheric neutrino anomalies can be explained 
by the transformation of electron and muon neutrinos into other neutrino flavors as they propagate from a source to a detector~\cite{Zyla:2020zbs:PDG2020Release}.

In the standard three-flavor neutrino framework, the neutrino flavor eigenstates (${\nu_e}$, ${\nu_\mu}$, ${\nu_\tau}$) are connected with the mass eigenstates ($\nu_1$, $\nu_2$, $\nu_3$) via the lepton flavor mixing matrix, also referred to as the Pontecorvo-Maki-Nakagawa-Sakata (PMNS)~\cite{Pontecorvo:1967fh,Maki:1962mu} matrix, $U_\text{\tiny PMNS}$
\begin{eqnarray}
\left(
\begin{matrix}
\nu_e  \cr
\nu_\mu  \cr
\nu_\tau  \cr
\end{matrix}
\right)
= 
U_\text{\tiny PMNS}
\left(
\begin{matrix}
\nu_1  \cr
\nu_2  \cr
\nu_3  \cr
\end{matrix}
\right)\,,
\end{eqnarray}
where the standard parametrization of $U_\text{\tiny PMNS}$ is given by~\cite{Zyla:2020zbs:PDG2020Release}
\begin{eqnarray}
   U_\text{\tiny PMNS} =
\left(
 \begin{matrix}
 1  &  0       & 0       \cr
 0  &  c_{23}  & s_{23}  \cr
 0  & -s_{23}  & c_{23}  \cr
\end{matrix}
\right)
\left(
\begin{matrix}
 c_{13} & 0  &  s_{13}e^{-i\delta_{\text{\tiny CP}}} \cr
 0  & 1 & 0 \cr -s_{13} e^{i\delta_{\text{\tiny CP}}} & 0 & c_{13} \cr
\end{matrix}
\right)
\left(
\begin{matrix}
c_{12} & s_{12} & 0 \cr
-s_{12} & c_{12} & 0 \cr
0  & 0 & 1 \cr
\end{matrix}
\right)
\left(
\begin{matrix}
e^{i\eta_1} & 0 & 0 \cr
0 & e^{i\eta_2}  & 0 \cr
0  & 0 & 1 \cr
\end{matrix}
\right),
\label{eq:U}
\end{eqnarray}
here the notations
$c_{ij} \equiv \cos \theta_{ij}$ and
$s_{ij} \equiv \sin \theta_{ij}$ are used. 
A nontrivial value of the Dirac CP phase, $\delta_{\text{\tiny CP}}$, results in the violation of the charge-conjugation-parity (CP) symmetry in neutrino oscillations.
Here, $\eta_i\;(i=1,2)$ are the Majorana CP phases that are physical only if neutrinos are Majorana-type particles and play no role in neutrino oscillations~\cite{Bilenky:1980cx}.

The calculation of neutrino oscillation probabilities also involves three mass squared differences defined as 
\begin{eqnarray}
\Delta m^2_{ij} \equiv m_i^2-m_j^2 \ \ \ (i,j = 1,2,3,\, i > j)\,, 
\label{eq:Dm2}
\end{eqnarray}
where $m_i$ is the rest mass of the mass eigenstate $\nu_i$. Note that two of the three  mass squared differences are independent in the three-flavor neutrino framework.
The three mixing angles $\theta_{12}$, $\theta_{23}$, $\theta_{13}$ and two independent mass squared differences $\Delta m^2_{21}$ and $|\Delta m^2_{31}|$ (or equivalently $|\Delta m^2_{32}|$) have been measured with a precision of a few percent~\cite{Zyla:2020zbs:PDG2020Release}. 
In the solar sector, $\Delta m^2_{21}$ is predominantly determined by the KamLAND reactor $\bar{\nu}_{e}$ disappearance experiment~\cite{KamLAND:2013rgu}, whereas $\theta_{12}$ is primarily measured by a combination of the solar neutrino experiments, such as SNO~\cite{SNO:2011hxd} and Super-Kamiokande~\cite{Super-Kamiokande:2016yck}.
In the atmospheric sector, $|\Delta m^2_{31}|$ and $\theta_{23}$ are primarily constrained by the ${\nu_\mu}$ and $\bar{\nu}_{\mu}$ disappearance accelerator and atmospheric neutrino experiments such as NO$\nu$A~\cite{NOvA:2021nfi}, T2K~\cite{T2K:2021xwb}, Super-Kamiokande~\cite{Super-Kamiokande:2019gzr}, and IceCube~\cite{IceCubeCollaboration:2023wtb}.
The mixing angle $\theta_{13}$, which connects the solar and atmospheric sectors, is best measured using the $\bar{\nu}_{e}$ disappearance reactor experiments, Daya Bay~\cite{DayaBay:2012fng,DayaBay:2022orm}, Double Chooz~\cite{DoubleChooz:2011ymz,DoubleChooz:2019qbj}, and RENO~\cite{RENO:2012mkc,RENO:2018dro}.
The sign of $\Delta m^2_{31}$ (or $\Delta m^2_{32}$), which is connected to the neutrino mass ordering (NMO) problem~\cite{Qian:2015waa}, and the value of the CP phase $\delta_{\text{\tiny CP}}$~\cite{Nunokawa:2007qh}, are the two major unknowns in the three-neutrino oscillation framework. 
Although some preliminary experimental hints exist~\cite{NOvA:2021nfi,T2K:2021xwb,Super-Kamiokande:2019gzr}, a conclusive determination must likely wait for the next generation of neutrino experiments. 

In literature, the neutrino mass eigenstates--$\nu_1$, $\nu_2$ and $\nu_3$ are defined with decreasing ${\nu_e}$ flavor content (i.e., $|U_{e1}|^2>|U_{e2}|^2>|U_{e3}|^2$)~\cite{deGouvea:2008nm}. 
$\Delta m^2_{21}>0$ is determined through the Mikheyev-Smirnov-Wolfenstein (MSW) matter effect~\cite{Wolfenstein:1977ue,Mikheyev:1985zog} of neutrino flavor conversion measured in solar neutrino experiments~\cite{Wurm:2017cmm}. 
Thus, whether $m_{1}<m_{2}<m_{3}$ (normal ordering, NO) or $m_{3}<m_{1}<m_{2}$ (inverted ordering, IO) is to be determined experimentally as both scenarios are compatible with the current data~\cite{Qian:2015waa}.
The determination of the NMO has significant implications in particle physics, nuclear physics, astrophysics and cosmology, including the origin of neutrino masses and flavor mixing, neutrinoless double beta decay searches~\cite{Dolinski:2019nrj}, supernova neutrino flavor conversion~\cite{Duan:2010bg}, nucleosynthesis~\cite{Kajino:2014bra}, and cosmological probes of the absolute neutrino mass scale~\cite{TopicalConvenersKNAbazajianJECarlstromATLee:2013bxd}.
Therefore, determination of the NMO is one of the most important tasks of the next generation neutrino oscillation experiments. 

Analogous to the determination of the sign of $\Delta m^2_{21}$, the determination of the NMO via the MSW matter effect is the foundation for long-baseline atmospheric and accelerator neutrino experiments, such as DUNE~\cite{DUNE:2021mtg}, Hyper-Kamiokande~\cite{Hyper-Kamiokande:2022smq}, PINGU~\cite{IceCube-PINGU:2014okk}, and ORCA~\cite{KM3NeT:2021ozk}. 
The Jiangmen Underground Neutrino Observatory (JUNO), with a baseline of 52.5~km, uses a different approach through the interference effect of the quasi-vacuum oscillation of reactor antineutrinos, which is unique and complementary to other neutrino oscillation experiments.

\begin{figure}
  \centering
  \includegraphics[width=0.9\textwidth]{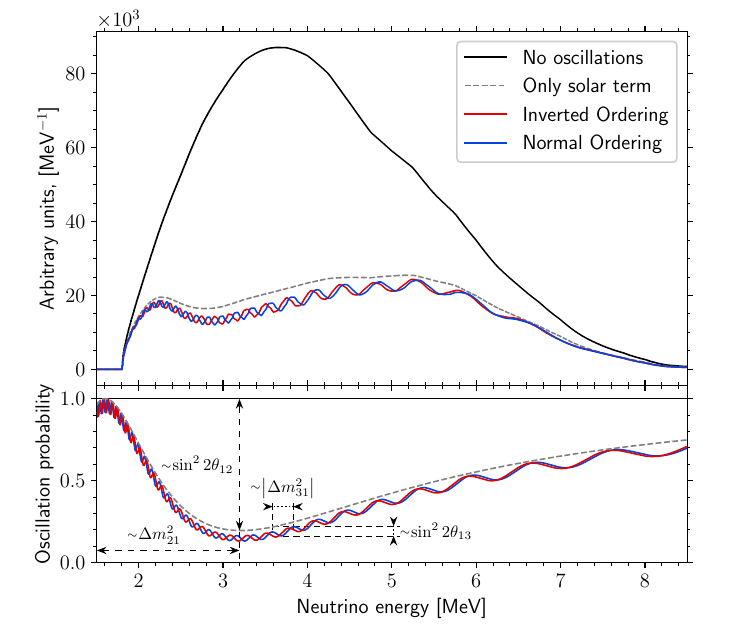}
  \caption{
  Expected antineutrino spectra at the JUNO detector under the assumption of perfect energy resolution (top) after 6 years of data taking with and without oscillations. 
  The blue and red lines indicate the normal and inverted ordering, respectively.
  The dotted line is scaled down by a factor 7 for better visibility. The survival probability is shown for the baseline of 52.5~km (bottom). 
  The solar term refers to a case in which only oscillations due to $\Delta m^2_{21}$ and $\sin^2\theta_{12}$ occur, i.e., $\sin^2\theta_{13} = 0$. 
  For this case, the value $\sin^2\theta_{12} = 0.282$ is used to avoid the overlap with red and blue curves, and other oscillation parameters are taken from PDG~2020~\cite{Zyla:2020zbs:PDG2020Release}.}
  \label{fig:oscillation:JUNO}
\end{figure}

The reactor antineutrino survival probability can be expressed as
\begin{eqnarray}
{\cal P}(\overline{\nu}^{}_e \to \overline{\nu}^{}_e)
&  =  & 
1 - \hspace{0.05cm} \sin^2 2\widetilde\theta_{12}\, {\tilde c}_{13}^4\, \sin^2 \widetilde\Delta_{21}
-  \sin^2 2\widetilde\theta_{13}
\left( \tilde c_{12}^2\sin^2 \widetilde\Delta_{31} + \tilde s_{12}^2 \sin^2 \widetilde\Delta_{32}\right) 
\nonumber \\
&  =  & 
1 - \hspace{0.05cm} \sin^2 2\widetilde\theta^{}_{12} \tilde c_{13}^4 \sin^2 \widetilde\Delta_{21}
- \frac{1}{2}
\sin^2 2\widetilde \theta^{}_{13} \left(\sin^2 \widetilde\Delta_{31}
+ \sin^2 \widetilde \Delta_{32}\right)   \hspace{0.5cm}
\nonumber \\
& & \hspace{0.3cm} - \hspace{0.1cm}
\frac{1}{2} \cos 2\widetilde \theta^{}_{12} \sin^2 2\widetilde \theta^{}_{13}
\sin \widetilde \Delta_{21}
\sin (\widetilde \Delta_{31} + \widetilde \Delta_{32}),
\label{eq:P_ee_mat}
\end{eqnarray}
where $\tilde c_{ij} \equiv \cos \widetilde \theta_{ij}$ and
$\tilde s_{ij} \equiv \sin \widetilde \theta_{ij}$,
with $\widetilde \theta_{ij}\, (i,j = 1,2,3,\, i<j)$ being the effective matter-induced mixing angles in $\widetilde U$ with the same parametrization as in Eq.~\eqref{eq:U};
$\widetilde\Delta_{ij} \equiv {\Delta \tilde m^2_{ij} L}/{4E}$, with $\Delta \tilde m^2_{ij}$ being the effective matter-induced mass squared difference;
$L$ is the baseline length; and $E$ is the neutrino energy.
In the second and third lines of Eq.~\eqref{eq:P_ee_mat},
we reformulate the survival probability to factor out the solar-dominated, atmospheric-dominated, and NMO-sensitive terms, respectively.
Note that the quantities marked with a tilde symbol represent the effective parameters in matter, serving as counterparts of their respective quantities in vacuum, which are obtained by the diagonalization of the effective Hamiltonian $\widetilde{\cal H}^{}_{\rm eff}$
\begin{eqnarray}
\widetilde{\cal H}^{}_{\rm eff}  = \frac{1}{2 E} \left[U
\begin{pmatrix} m^2_1 & 0 & 0 \cr 0 & m^2_2 & 0 \cr 0 & 0 & m^2_3
\cr \end{pmatrix} U^\dagger - \begin{pmatrix} A & 0 & 0 \cr 0 & 0 &
0 \cr 0 & 0 & 0 \cr \end{pmatrix} \right] = \frac{1}{2 E} \left[\widetilde{U}
\begin{pmatrix} \widetilde{m}^2_1 & 0 & 0 \cr 0 & \widetilde{m}^2_2
& 0 \cr 0 & 0 & \widetilde{m}^2_3 \cr \end{pmatrix}
\widetilde{U}^\dagger \right]\; ,
\end{eqnarray}
in which the matter potential $A$ is derived as~\cite{Wolfenstein:1977ue,Mikheyev:1985zog}
\begin{eqnarray}
  A=2\sqrt{2} \ G^{}_{\rm F} N^{}_e E\simeq 1.52 \times 10^{-4} ~{\rm eV}^2 \cdot Y_e^{}
 \cdot \frac{\rho}{{\rm g}/{\rm cm}^{3}} \cdot \frac{E}{\rm GeV},
\end{eqnarray}
here $G^{}_{\rm F}$ is the so-called Fermi constant, $N^{}_e$ is the number density of electrons, $Y_e^{} \simeq 0.5$ is the electron fraction per nucleon, and $\rho = (2.45\pm0.15) ~{\rm g}/{\rm cm}^3$ is the estimated average matter density with its associated uncertainty~\cite{JUNO:2022mxj}.
Both the exact calculation and analytical approximations of the effective mass and mixing parameters in Eq.~\eqref{eq:P_ee_mat} have been employed, and consistent results have been obtained.
Note that matter corrections to the $\bar{\nu}_{e}$ survival probability (e.g., around the $1\%$ level for $\theta_{12}$ and $\Delta m^2_{21}$) cause a slight degradation in the NMO sensitivity and are essential in the precision measurement of oscillation parameters~\cite{Li:2016txk,Capozzi:2013psa}.
The reactor electron antineutrino survival probability at the JUNO baseline of 52.5~km is shown in~Figure~\ref{fig:oscillation:JUNO} (bottom). 
For illustration, the impact of oscillations on the antineutrino spectrum expected in JUNO under the assumption of perfect energy resolution is shown on top (more details are given in Section~3).
The values of the neutrino oscillation parameters are taken from the 2020 release of the Particle Data Group (PDG~2020)~\cite{Zyla:2020zbs:PDG2020Release}: $\sin^2\theta_{12}=0.307\pm 0.013$, $\Delta m_{21}^2=(7.53\pm 0.18)\times 10^{-5}\ {\rm eV}^2$, $\Delta m_{32}^2=(-2.546^{+0.034}_{-0.040} )\times 10^{-3}\ {\rm eV}^2$ (IO), $\Delta m_{32}^2=(2.453\pm 0.034)\times 10^{-3}\ {\rm eV}^2$ (NO), and $\sin^2\theta_{13}=(2.18\pm 0.07)\times 10^{-2}$.
Figure~\ref{fig:oscillation:JUNO} clearly shows that a spectral measurement with high energy resolution is a prerequisite for determining the NMO with reactor antineutrinos. 

JUNO was designed primarily for this purpose using a 20 kton liquid scintillator (LS) detector located approximately 52.5 km from the Taishan and Yangjiang nuclear power plants (NPPs) in Jiangmen City, Guangdong Province, China.
In this study, we update the NMO sensitivity in JUNO from the previous estimate given in Ref.~\cite{An:2015jdp}. 
Only eight nuclear reactors are considered at the distance of approximately 52.5 km instead of the ten assumed in Ref.~\cite{An:2015jdp}, when the JUNO experiment was conceived.
Considering the latest detector geometry, the increase in the photomultiplier tube (PMT) detection efficiency, and the latest PMT optical model~\cite{Wang:2022tij}, an improved energy resolution model~\cite{JUNO:EnergyResolutionPaper} was developed based on the latest official simulation framework and is used here.
For the first time, we perform a joint analysis of JUNO and its satellite experiment Taishan Antineutrino Observatory (TAO)~\cite{JUNO:2020ijm}, which provides a constraint of the reactor antineutrino spectral shape.
Other updates include the actual location and overburden of the experimental site, new inverse beta decay (IBD) selection and muon veto efficiencies estimated using a full simulation, the consideration of the atmospheric neutrino-induced neutral current background and antineutrino flux from the remote world reactors, and consideration of terrestrial matter effects. 
All these new inputs are described in the following sections.
This paper focuses on JUNO's sensitivity to the NMO using reactor antineutrinos exclusively. 
However, note that incorporating external constraints on $\Delta m_{32}^2$ (or $\Delta m_{31}^2$) from atmospheric and accelerator neutrino experiments, as well as JUNO's own atmospheric neutrino measurements, can enhance the NMO sensitivity~\cite{Li:2013zyd,An:2015jdp, Nunokawa:2005nx}.

The remainder of this paper is organised as follows. In Section~\ref{sec:response}, we introduce the JUNO detector, focusing on the detector response model used in this analysis. 
In Section~\ref{sec:signal_background}, we calculate the IBD signal and the corresponding backgrounds. The experimental setup of TAO and the expected signal and backgrounds are presented in Section~\ref{sec:tao}.
Section~\ref{sec:analysis_results} describes the methodology used to evaluate the NMO sensitivity and the main results and discussions. Section~\ref{sec:conclusion} provides the conclusions.

\section{JUNO Detector}
\label{sec:response}

The JUNO detector~\cite{An:2015jdp, JUNO:2022hxd} is under construction in an underground laboratory under Dashi Hill in Guangdong Province, South China. 
A 650~m rock overburden (1800~m.w.e) can effectively suppress the cosmic muon flux to $ 4.1 \times  10^{-3}/(\mathrm{s} \cdot \mathrm{m}^2)$.
A 20 kton LS~\cite{JUNO:2020bcl} target, consisting of $\sim$88\% carbon and $\sim$12\% hydrogen by mass, is contained in a 12-cm thick acrylic sphere with a 35.4 m inner diameter.
To determine the NMO by precisely measuring the neutrino oscillation pattern, an energy resolution better than 3\% at 1 MeV should be achieved, which is unprecedented for a detector of this type.
The acrylic sphere is surrounded by 17,612 large 20-inch high quantum efficiency PMTs, referred to as LPMTs, and 25,600 small 3-inch PMTs, referred to as SPMTs, yielding a total photo-cathode coverage of 78\%. 
An ultra-pure water buffer with a minimal thickness of $\sim$1.5~m fills the volume between the acrylic sphere and LPMT photocathode. 
This center detector (CD) is fully surrounded by an ultra-pure water Cherenkov detector that serves as an active
veto for cosmic muons and as a passive shield against external radioactivity and neutrons from cosmic rays. 
The minimal thickness of the water detector is 2.5~m.  
The cosmic muon veto system is supplemented with an external muon tracker consisting of three layers of plastic scintillator refurbished from the OPERA experiment~\cite{Acquafredda:2009zz} located at the top, providing
a muon track angular reconstruction precision of 0.20$^{\circ}$~\cite{JUNO:2023cbw}. 
This system~\cite{JUNO:2023cbw} covers approximately 60\% of the surface above the water pool with the primary objective of providing a sample of well reconstructed muons that can be used to benchmark the reconstruction of the water pool and CD.
For more information about the JUNO detector, please refer to Ref.~\cite{JUNO:2022hxd}.

JUNO detects reactor antineutrinos via the IBD reaction on hydrogen in the LS, 
$\bar{\nu}_e + p \rightarrow e^{+} + n$. 
The positron quickly deposits its kinetic energy in the LS and annihilates into two 0.511~MeV gammas, generating optical photons through the scintillation and Cherenkov processes that are then detected by the PMTs. 
This forms a prompt signal whose energy is typically deposited a few nanoseconds after the IBD reaction.
After scattering in the LS with an average lifetime of approximately 200~\textmu s,
the neutron is captured by a hydrogen or carbon nucleus with roughly 99\% and 1\% probabilities, consequently producing a delayed gamma signal of 2.22~MeV or 4.95~MeV, respectively.
The energy signature of the prompt and delayed signals, as well as their temporal and spatial correlation, constitute effective handles for separating the IBD signal from the backgrounds.

Precise measurement of the antineutrino energy is essential for determining the NMO via the spectral distortion caused by the neutrino oscillation pattern shown in Figure~\ref{fig:oscillation:JUNO}.
Dedicated energy reconstruction algorithms have been developed to precisely determine the IBD prompt energy~\cite{Huang:2022zum}.
A detector energy response model is constructed as a matrix, $R(E^{\nu}, E^{\mathrm{rec}})$, mapping the antineutrino energy ($E^{\nu}$) to the reconstructed energy of the IBD prompt signal ($E^{\mathrm{rec}}$).
The matrix is constructed by applying three effects in sequence to the simulated antineutrinos of defined energies: kinematics of the IBD reaction, energy nonlinearity due to the scintillation and Cherenkov processes, and energy resolution.
Energy leakage caused by escaping secondary particles affects less than 1\% of the IBD events owing to the application of a fiducial volume cut; thus, energy leakage is not considered, resulting in a negligible impact on the energy response.

\subsection{IBD Reaction Kinematics}
In the IBD reaction, the electron antineutrino transfers most of its energy to the positron.
The deposited energy (\Edep) for the IBD prompt signal is defined to include the positron kinetic energy and annihilation energies of the two 0.511~MeV $\gamma$s.
The energy threshold of the IBD reaction is 1.8~MeV. 
Consequently, \Edep is approximately equal to $E^{\nu} - 0.78$~MeV, which considers the annihilation energy.
More precisely, the kinetic energy of the positron also depends on the scattering angle of the positron with respect to the incident antineutrino.
This results in an energy spread of the positron even for a fixed energy of incident antineutrinos. 
The energy spread causes a nontrivial effect on the deposited energy distribution.
Figure~\ref{fig:reponese:pdfs} shows the distributions of the deposited energy, which resemble rectangular distributions with sloping tops, for incident antineutrino energies of 3~MeV, 4~MeV, and 5~MeV, respectively. 
We integrate the scattering angle using the double differential IBD cross section, which is a function of the neutrino energy and scattering angle~\cite{Strumia:2003zx}. 
The final-state neutron in the IBD reaction carries a few tens of keV of kinetic energy, which is anti-correlated to the energy spread of the positron. 
The neutron kinetic energy is mostly undetectable in the detector owing to the large scintillation quenching of any protons or carbon nuclei it might scatter with.
By adding the kinetic energy of the neutron after quenching to the prompt energy of the IBD process, the positron energy spread would be partially cancelled.
In this analysis, we ignore the neutron kinetic energy and consider only the positron kinetic energy and its spread.

\subsection{Nonlinear Energy Response}
\label{subsec:lsnl}
The relation between the deposited energy and number of scintillation photons detected by PMTs is not linear, primarily owing to quenching.
The Cherenkov process contributes $<10\%$ photons in the energy region of IBD positron in JUNO LS.
The Cherenkov radiation depends on the particle's track length above the Cherenkov threshold and is also a nonlinear function of the positron kinetic energy.
The instrumental charge nonlinearity of the JUNO PMTs and the associated electronics can be calibrated with a residual less than 0.3\% owing to the dual-calorimetry calibration technique~\cite{JUNO:2020xtj}.
The energy nonlinearity distorts the prompt energy spectrum and is thus essential in determining the NMO.

We define the liquid scintillator nonlinearity (LSNL) as $E^{\mathrm{vis}}/E^{\mathrm{dep}}$, where $E^{\mathrm{dep}}$ is the deposited energy in the scintillator, and $E^{\mathrm{vis}}$ is the visible energy defined as the expected reconstructed energy assuming perfect energy resolution.
The LSNL is different between positrons, i.e., signal events, and
gammas, i.e., calibration events, because of their different energy deposition patterns.
In this paper, the term LSNL refers to the positron LSNL by default, which is primarily calibrated using gamma calibration sources, considering the conversion between positron and gamma.
The nonlinearity response of the JUNO detector will be measured with similar calibration sources and procedures as in Daya Bay~\cite{Adey:2019zfo:DYB_NL}, which has a similar scintillator composition.
Therefore, in this sensitivity study, we assume that a comparable precision can be reached and the relative systematic uncertainty of the nonlinearity calibration is the same as in Daya Bay~\cite{Adey:2019zfo:DYB_NL}.
A nominal nonlinearity curve shown in the inset of Figure~\ref{fig:reponese:pdfs} is obtained through a simulation of the IBD positrons in the JUNO detector, including all detector effects.
The nominal nonlinearity curve is used in calculating of the visible energy spectrum. 
The methodology employed in the Daya Bay experiment~\cite{Adey:2019zfo:DYB_NL} is adopted for JUNO to account for the systematic uncertainty of nonlinearity.
A Monte Carlo method is used to generate various nonlinearity curves by sampling and fitting the calibration data while considering their uncertainties.
Among these various nonlinearity curves, four basic curves are selected to represent four typical variations with one standard deviation.
A variable nonlinearity curve is generated by the combination of the nominal nonlinearity curve and the four basic nonlinearity curves.
The weighting factors used in the combination are assigned as four constrained nuisance parameters, which effectively control the variations in the prediction of the energy spectrum.

\subsection{Energy Resolution}
An excellent energy resolution is required to effectively discriminate between the two possible NMO signatures in the prompt energy spectrum.
Therefore, understanding the different elements that contribute to the energy resolution is critical for this study.  
The detector's energy resolution is dominated by the statistical fluctuation of the number of detected photoelectrons (PE).
To consider the systematic effects in the energy resolution, we define a parametrization formula for the relative energy resolution of the prompt signal:
\begin{equation}
    \frac{\sigma_{E^{\rm rec}}}{E^{\rm vis}}=\sqrt{\left(\frac{a}{\sqrt{E^{\rm vis}}}\right)^2+b^2+\left(\frac{c}{E^{\rm vis}}\right)^2},
    \label{eq:EnergyResolution:abc}
\end{equation}
where $a$ is the term driven by the Poisson statistical fluctuation of the number of detected photoelectrons,
$b$ is dominated by residual effects after correcting the detector’s spatial non-uniformity, and $c$ is dominated by the PMT dark noise and the fluctuations in energy deposition for the annihilation gammas.
Other systematic effects contributing to the energy resolution, which are of minor importance, are also considered in this formula~\cite{JUNO:EnergyResolutionPaper}.

\begin{figure}[htb]
    \centering
    \includegraphics[width=0.9\textwidth]{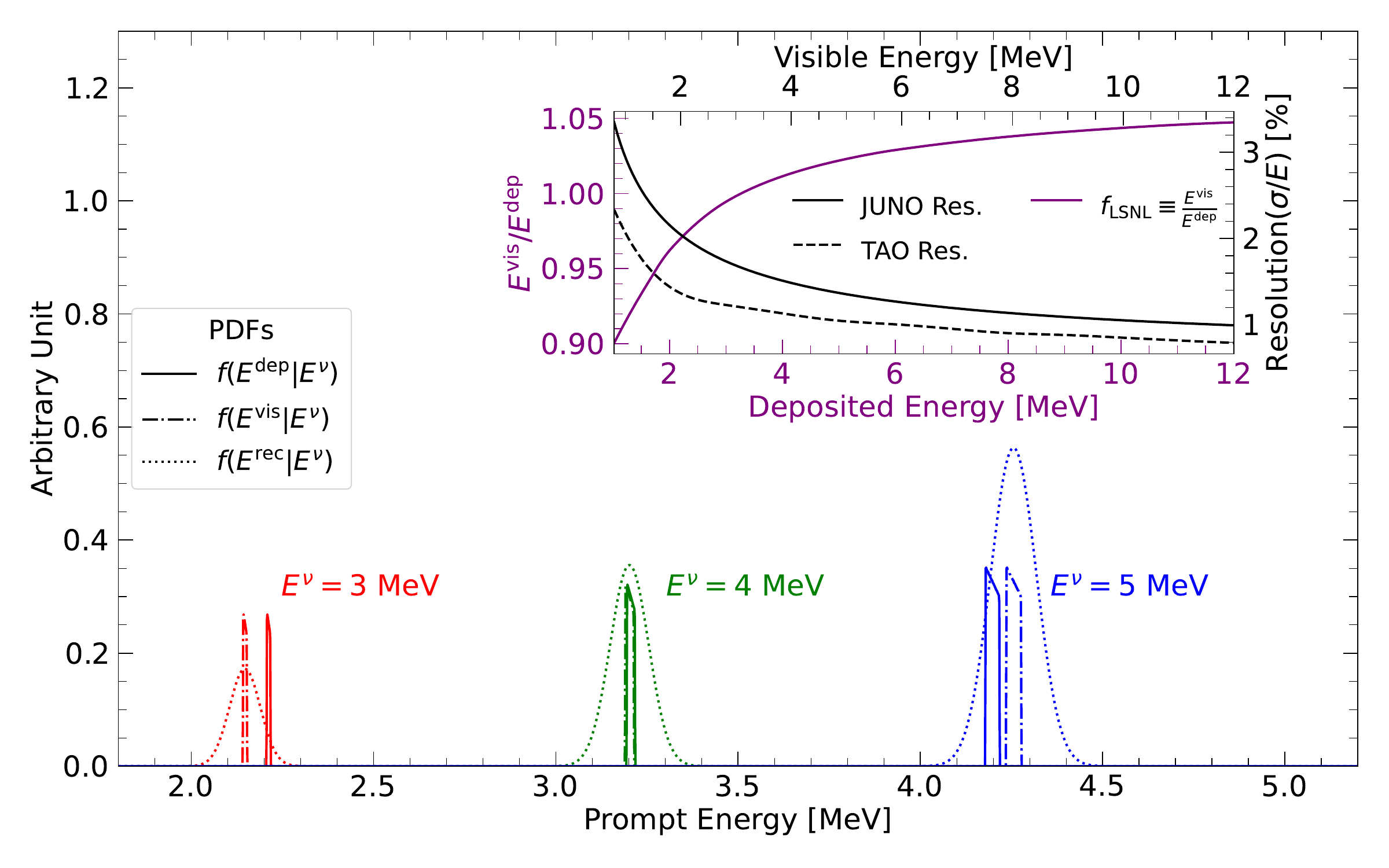}
    \caption{Energy conversion probability density functions (PDFs)  of the deposited, visible, and reconstructed energies of the IBD prompt signal for three typical antineutrino energies. 
    The deposited energy, which is approximately equal to $E^{\nu} - 0.78$~MeV, includes the positron kinetic and annihilation energies. 
    The conversion from the deposited energy to visible energy (proportional to the expected number of photons) is described by the positron liquid scintillator nonlinearity (LSNL), and the conversion to the reconstructed energy is described with the energy resolution. 
    The areas of the PDFs are proportional to the cross sections at the  three typical antineutrino energies.
    The LSNL curve, which is the same for both JUNO and TAO, is shown in the inset.
    The energy resolutions for JUNO and TAO are also shown in the inset.
    }
    \label{fig:reponese:pdfs}
\end{figure}

A previous study showed that JUNO could achieve a photoelectron yield of 1345~PE/MeV for the neutron capture on hydrogen at the detector center and an energy resolution of 3\% at 1~MeV in the fiducial volume\cite{JUNO:2020xtj}.
In this analysis, the photoelectron yield is updated to be 1665~PE/MeV, and the energy resolution is improved to be $2.95\%$ at 1~MeV. 
This improvement is a result of a deeper understanding of the energy resolution using a full simulation and the latest information available on the detector design and construction, including photoelectron statistics, scintillation quenching, Cherenkov radiation, dark noise,  charge response of the PMTs, readout electronics, vertex reconstruction resolution, and non-uniformity in energy reconstruction. 
The details of this study are available in Ref.~\cite{JUNO:EnergyResolutionPaper}, and the following are the major updates contributing to the improvement:
\begin{itemize}
    \item The average photon detection efficiency of the CD PMTs is updated to 30.1\% based on the mass testing data of PMTs~\cite{JUNO:2022hlz}, which is better than the nominal requirement of 27\%. Consequently, the photoelectron yield is increased by a relative 11\%.
  \item The latest detector geometry in the detector simulation increases the photon acceptance with better modelling of the reflective surfaces of the detector components, resulting in a 3\% increase in the photoelectron yield.
  \item With the measured angular and spectral dependencies of the PMTs’ detection efficiency, a unified optical model is implemented in the detector simulations of Daya Bay and JUNO that accounts for photon interactions in the glass of the PMTs and the optical processes occurring inside the PMT volume~\cite{Wang:2022tij}. 
  The photoelectron yield increases by 8\% after the JUNO simulation result is benchmarked with the photoelectron yield observed in the Daya Bay experiment~\cite{JUNO:EnergyResolutionPaper}.
\end{itemize}
The updated energy resolution curve is shown in the inset of Figure~\ref{fig:reponese:pdfs} as a function of the reconstructed IBD prompt energy for events uniformly distributed in the fiducial volume.
We simulate many positrons at various fixed energies and fit their energy distributions with Gaussian functions to obtain their energy resolutions. 
By fitting the obtained energy resolutions with the expression given in Eq.~\eqref{eq:EnergyResolution:abc}, we find that $a = 2.61\%~\sqrt{\rm MeV}$, $b$ = 0.64\%, and $c = 1.20\%~{\rm MeV}$.
The energy resolution uncertainty depends on the calibration strategy performed in the JUNO detector.
A Monte Carlo simulation study of various calibration sources and the extraction of energy resolution were also reported in Ref.~\cite{JUNO:EnergyResolutionPaper}.  
The absolute uncertainties on $a$, $b$, and $c$ are $0.02\%~\sqrt{\rm MeV}$, 0.01\%, and 0.04\%~MeV, respectively, and are used in this analysis. 

Figure~\ref{fig:reponese:pdfs} shows the deposited, visible, and reconstructed energy distributions for reactor antineutrinos of three different energies.
The reconstructed energy distributions are slices of the full energy response matrix.
The nominal nonlinearity and energy resolution of JUNO are shown in the inset of Figure~\ref{fig:reponese:pdfs} as functions of deposited and visible energy.
A similar procedure is followed for the TAO detector response, as described in Section~\ref{sec:tao}.
The TAO detector has a better energy resolution, which is also shown in the inset of Figure~\ref{fig:reponese:pdfs}.
The visible energy scale is determined by the choice of the n-H capture gamma peak at 2.2 MeV as the anchor point, which forces the gamma $E^{\rm vis}/E^{\rm dep}$ nonlinearity curve (not shown) to cross 1 at 2.2 MeV. 
At this energy scale, the positron nonlinearity curve crosses 1 for deposited energy at approximately 3.2~MeV.
The visible energy shift with respect to the deposited energy is due to the LSNL.

\section{Expected Signal and Backgrounds} 
\label{sec:signal_background}

\subsection{Reactor Antineutrino Flux}
\label{sec:flux}
The reactor antineutrinos detected at JUNO primarily originate from the Taishan and Yangjiang NPPs, which have two and six reactor cores, respectively.
The location of the JUNO experiment is selected to be almost equidistant from all the reactor cores.
Information on the real-time operation of the reactors in the Taishan, Yangjiang, and Daya Bay NPPs is essential to predict the reactor antineutrino flux and spectrum at JUNO.
The reactor power, baselines, and expected IBD rates after oscillation from the Taishan, Yangjiang, and Daya Bay reactor cores are summarized in  Table~\ref{tab:Reactor}.
These were also used as inputs for the study in Ref.~\cite{JUNO:2022mxj}, which presented an updated sensitivity to the oscillation parameters.
The antineutrinos from Daya Bay carry essentially no information of the NMO because of the difference in baseline from that of the Taishan and Yangjiang NPPs, and these antineutrinos slightly reduce the NMO sensitivity.
The Huizhou NPP, located 265~km away, will not begin operation until several years after data taking and it has not been considered in this analysis owing to its small impact and uncertain schedule. 
The other running NPPs are more than 300~km away, and their contribution is estimated to be approximately one IBD event per day according to the world nuclear reactor information \cite{international2019iaea}.
In this study, they are treated as the ``world reactors'' background since their signals provide no useful information on the NMO.

\begin{table}[htb]
\begin{center}
\begin{tabular}{lcccc}
    \hline
    Reactor   & Power [GW$_{\rm th}$] & Baseline [km] & IBD Rate [day$^{-1}$]  \\
    \hline
    Taishan & 9.2   & 52.71 & 18.4    \\
    ~~~Core 1 & ~~~4.6   & ~~~52.77 & ~~~9.2    \\
    ~~~Core 2 & ~~~4.6   & ~~~52.64 & ~~~9.2    \\
    Yangjiang & 17.4  & 52.46 & 35.3    \\
    ~~~Core 1 & ~~~2.9   & ~~~52.74 & ~~~5.8    \\
    ~~~Core 2 & ~~~2.9   & ~~~52.82 & ~~~5.8    \\
    ~~~Core 3 & ~~~2.9   & ~~~52.41 & ~~~5.9    \\
    ~~~Core 4 & ~~~2.9   & ~~~52.49 & ~~~5.9    \\
    ~~~Core 5 & ~~~2.9   & ~~~52.11 & ~~~6.0    \\
    ~~~Core 6 & ~~~2.9   & ~~~52.19 & ~~~5.9    \\
    Daya Bay & 17.4  & 215  & 3.65   \\ 
    Total IBD rate &  &  & 57.4 \\ \hline
\end{tabular}
\caption{\small {Characteristics of NPPs and their reactor cores considered as antineutrino sources in this analysis:  Taishan and Yangjiang, at an average distance of 52.5~km, and the next closest, Daya Bay. 
The IBD rates are estimated from the baselines, full thermal power, and oscillation probability with the neutrino oscillation parameters listed in Section~\ref{sec:intro}. 
}}
\label{tab:Reactor}%
\end{center}
\end{table}%

The expected reconstructed energy spectrum of the IBD signal observed at the JUNO detector can be expressed as
\begin{equation}
  \label{eq:energy_response}
\begin{split}
    S(E^{\rm rec}) = N_p\cdot\epsilon\cdot\int\limits_{T_{\rm DAQ}}dt
    \int\limits_{1.8~\mathrm{MeV}}^{15~\mathrm{MeV}}dE^{\nu}
     \cdot \Phi(E^{\nu},t)\cdot \sigma(E^{\nu})
    \cdot R(E^{\nu}, E^{\rm rec}),
\end{split}
\end{equation}
where $N_p = 1.44\times 10^{33}$ is the number of free protons in the 20-kton LS used as the detector target, $\epsilon = 82.2\%$ is the IBD event selection efficiency to be introduced in Section~\ref{sec:selection}, $T_{\mathrm{DAQ}}$ is the data-taking time, $\Phi(E^{\nu},t)$ is the oscillated reactor antineutrino flux in JUNO at time $t$,  $\sigma(E^{\nu})$ is the IBD cross section, and $R(E^{\nu}, E^{\rm rec})$ is the detector energy response function that includes the effects described in Section~\ref{sec:response}. 
The integration interval of the antineutrino energy begins from the IBD reaction threshold at 1.8 MeV and ends at 15~MeV, above which the reactor antineutrino flux becomes negligible.

In commercial reactors, electron antineutrinos are generated from the decays of the fission products of four major isotopes, $^{235}$U, $^{238}$U, $^{239}$Pu, and $^{241}$Pu, which contribute more than $99\%$ of the total neutrino flux above 1.8~MeV. 
The oscillated antineutrino flux at time $t$ is predicted as
\begin{equation}\label{equ_generic}
    \Phi(E^{\nu}, t)  = \sum_r \frac{{\cal P}_{\bar{\nu}_{e} \to \bar{\nu}^{}_e}(E^{\nu},L_{r})}{ 4\pi L^2_{r}}\frac{W_r(t)}{\sum_i f_{ir}(t) e_i}\sum_i f_{ir}(t)s_i(E^{\nu}),
\end{equation}
where ${\cal P}_{\bar{\nu}_{e} \to \bar{\nu}^{}_e}(E^{\nu},L_{r})$ is the antineutrino survival probability at a distance of $L_r$ from reactor $r$, with $r$ running over the reactor cores; $W_r(t)$ is the reactor thermal power; $f_{ir}(t)$ is the fission fraction of isotope $i$ among the four; $e_i$ is the mean energy released per fission for isotope $i$; and $s_i(E^{\nu})$ is the antineutrino energy spectrum per fission for each isotope.
The reactor thermal power and fission fractions are time-dependent and will be provided by the NPPs during the data-taking period.
In this study, the reactor thermal power and fission fractions are assumed to be stable at their averaged values without loss of generality.
To account for the loss of time incurred by refueling the reactors, which typically requires one month per year, we use a duty cycle factor of 11/12 to scale down the thermal power.
The average fission fractions are 0.58, 0.07, 0.30, and 0.05, with mean energies of 202.36~MeV, 205.99~MeV, 211.12~MeV, 214.26~MeV~\cite{Ma:2012bm} released per fission for $^{235}$U, $^{238}$U, $^{239}$Pu, and $^{241}$Pu, respectively.

The $\bar\nu_e$ energy spectra per fission of $^{235}$U, $^{238}$U, $^{239}$Pu, and $^{241}$Pu from the Huber-Mueller model~\cite{Huber:2011wv, Mueller:2011nm} are widely used to predict the reactor antineutrino flux.
However, the total observed antineutrino yield per fission exhibits a $\sim$5\% deficit compared with the model's prediction, which is referred to as the reactor antineutrino anomaly~\cite{Mention:2011rk}.
Furthermore, recent reactor antineutrino experiments including Daya Bay~\cite{An:2015nua}, RENO~\cite{RENO:2018dro}, Double Chooz~\cite{DoubleChooz:2019qbj}, NEOS~\cite{Ko:2016owz}, PROSPECT~\cite{PROSPECT:2022wlf}, STEREO~\cite{STEREO:2022nzk}, and DANSS~\cite{Danilov:2022bss} have observed some discrepancies between the measured and predicted antineutrino spectral shape, most notably in the form of an excess in the data at approximately 5 MeV.
The reactor antineutrino rate and spectrum were precisely measured in the Daya Bay experiment, and the spectral ratio to the Huber-Mueller model is presented in Ref.~\cite{An:2016srz}.
In this analysis, the $\bar\nu_e$ energy spectra per fission from the Huber-Mueller model corrected by the measurements in the Daya Bay experiment~\cite{An:2016srz} are used as initial values in the spectral fit.

In the measured beta decay spectra that are used as a reference in the Huber-Mueller model, the beta decay rates of some long-lived fission fragments do not reach equilibrium owing to limited time, whereas, in the reactor core, these fission fragments accumulate and reach equilibrium.
Therefore, this so-called ``non-equilibrium'' effect contributes an additional 0.6\% antineutrino flux based on the evaluation in Ref.~\cite{Mueller:2011nm}.
The spent nuclear fuel removed from the reactor cores during refueling and placed in cooling pools nearby still emits antineutrinos and contributes an additional 0.3\% antineutrino flux based on the calculations in Ref.~\cite{An:2016srz}.
Additional corrections for the antineutrino flux rate and spectrum are applied to account for these two effects.
Both the non-equilibrium and spent nuclear fuel contributions are assigned a 30\% rate uncertainty and a negligible spectrum shape uncertainty, as motivated from the experience of Daya Bay~\cite{An:2016srz}.
The realistic reactor electron antineutrino spectrum expected at the JUNO detector as a function of neutrino energy with and without oscillations is shown in Figure~\ref{fig:oscillation:JUNO}. 

The uncertainties in the reactor antineutrino flux are propagated to the predicted IBD event rate.
The baselines can be measured to a 1~m precision, resulting in a negligible uncertainty at JUNO baselines.
The reactor power will be measured and provided by the NPPs with an uncorrelated uncertainty of 0.5\% based on the experience of Daya Bay~\cite{An:2016srz}.
Similarly, the fission fractions will also be provided with an uncertainty of 5\% for the four major isotopes.
The uncertainties of the fission fractions are partially correlated~\cite{Ma:2014bpa} among the four isotopes and will contribute an uncertainty of 0.6\% to the predicted IBD rate~\cite{An:2016srz}. 
Both the uncertainties of reactor power and fission fractions are treated as uncorrelated between reactor cores as they are measured individually.
The mean energy per fission contributes only a 0.2\% uncertainty~\cite{Ma:2012bm}.
Finally, a 2\% correlated uncertainty is assigned to the IBD yield per fission, which is the product of the IBD cross-section with the antineutrino spectrum.
All these uncertainties are obtained directly from the experience of the Daya Bay experiment~\cite{An:2016srz}.

The uncertainty in the reactor antineutrino flux spectral shape is constrained by using the measurements from TAO, which will be a satellite detector with the primary goal of providing a precise model-independent reference spectrum for JUNO~\cite{JUNO:2020ijm}. 
More details about the TAO detector are given in Section~\ref{sec:tao}.
In this analysis, two methods are used to incorporate the constraints from TAO.
The first method performs a joint fit of JUNO and TAO data.
The detector response, expected IBD signal and backgrounds, and their uncertainties are implemented in the analysis for TAO using a similar procedure as in JUNO. 
In this analysis, JUNO and TAO share the same antineutrino energy spectrum before oscillation.
The reactor antineutrino spectral shape is expressed using a set of free parameters in the spectral fit to avoid any dependency on the reactor antineutrino flux model.
The second method is inspired by the recent study quantifying JUNO’s sensitivity to the oscillation parameters~\cite{JUNO:2022mxj}, where the expected uncertainty from TAO’s measured spectrum is assigned as a flux spectral shape uncertainty for JUNO.
Both methods produce consistent results. 
The results of the first method are reported as nominal, whereas those of the second are treated as a cross-check.

\subsection{IBD Event Selection}
\label{sec:selection}
Significant efforts have been made in the design and construction of the JUNO experiment to suppress backgrounds applying various strategies, including radioactivity screening and control of the detector material~\cite{JUNO:2021kxb}, overburden of the experimental hall to reject cosmic muons, and sufficient shielding.  
Furthermore, the unique prompt-delayed temporal and spatial coincidence signature of IBD events is an effective handle for rejecting background events.
A fiducial volume cut on the LS target is applied to further suppress backgrounds due to the natural radioactivity from the PMTs and other materials outside the LS region.  
Cosmic muon veto cuts are applied to suppress the cosmogenic backgrounds generated by muons passing through the detector.
Consequently, the IBD selection significantly reduces the background-to-signal ratio, from the level of $10^6$ in the raw experimental data, to less than 0.1 in the final selected IBD candidates. 

\begin{figure}[tb!]
    \centering
    \includegraphics[width=0.9\textwidth]{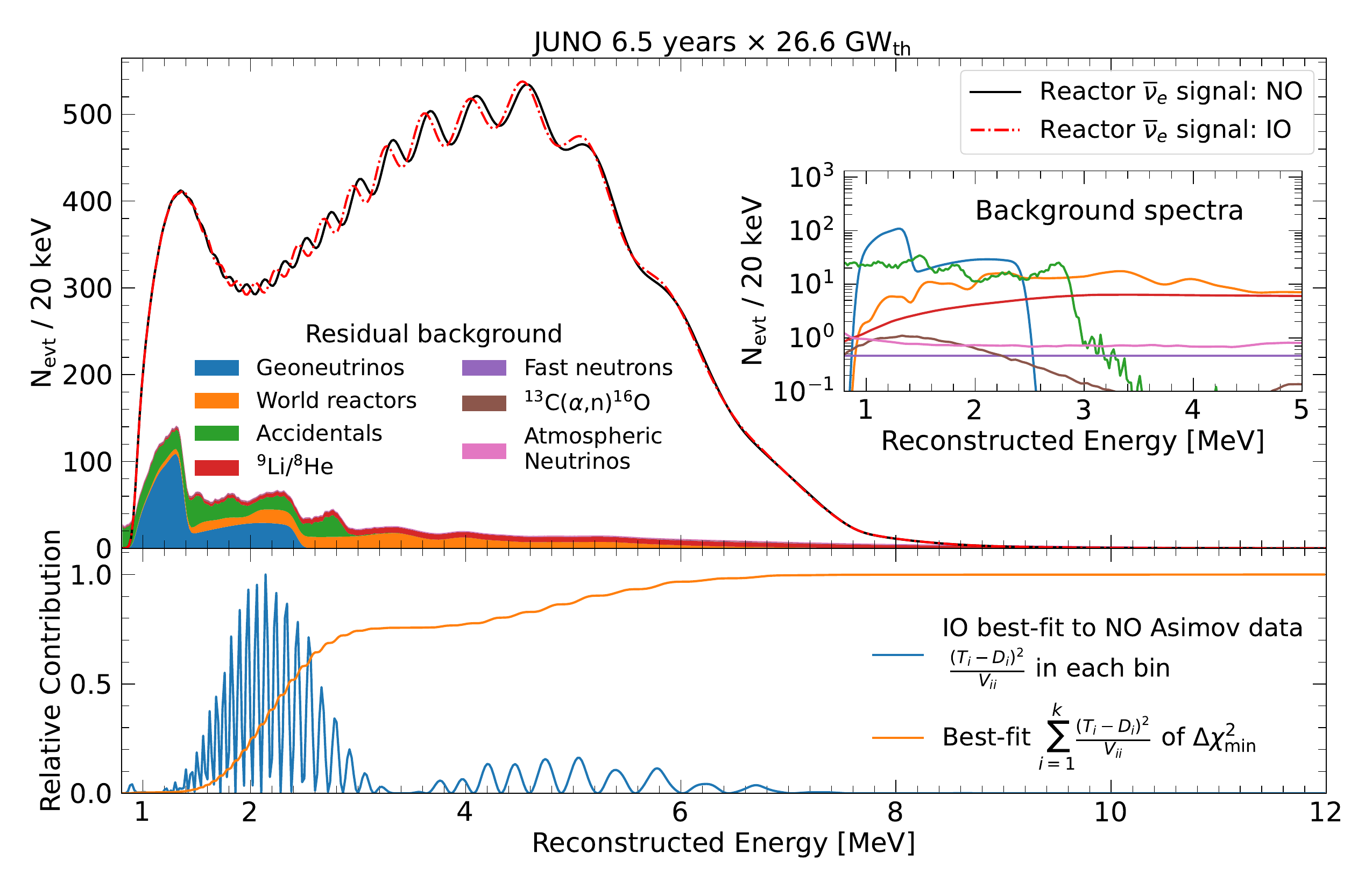}
    \caption{Top: Reconstructed energy spectra of JUNO in both the NO and IO scenarios without any statistical or systematic fluctuations (Asimov data). The neutrino oscillation parameters from PDG~2020~\cite{Zyla:2020zbs:PDG2020Release} are used to calculate the oscillation probability. 
    The background spectra in the main figure are stacked, whereas those in the inset are plotted individually.
    Bottom: Relative contribution to $\Delta\chi^2$ and cumulative $\Delta\chi^2$ obtained when fitting the IO spectrum with the NO hypothesis. 
    The best fit value of $\Delta m^2_{32}$ in the IO spectrum differ from the values used as inputs in the top panel.  
    The results show that the most sensitive region for JUNO's NMO determination is 1.5--3~MeV.
    }
    \label{fig:JUNO:signal_and_background}
\end{figure}

A new IBD selection with an optimized veto strategy has been developed both for this study on NMO sensitivity and a study on oscillation parameter precision in Ref.~\cite{JUNO:2022mxj}.
As shown in Figure~\ref{fig:JUNO:signal_and_background}, most of prompt energies of IBD signals are within an energy window of 1~MeV to 8~MeV. 
The prompt IBD candidate events are restricted to the energy window $[0.8, 12.0]$~MeV.
The delayed IBD candidate events are dominated by neutrons captured on hydrogen and carbon, releasing gammas with total energies of 2.2~MeV and 4.9~MeV, respectively. 
Correspondingly, the delayed energy windows are selected to be either $[1.9, 2.5]$~MeV or $[4.4, 5.5]$~MeV.
The IBD candidate events are discarded if the reconstructed positions of their prompt or delayed events are more than 17.2~m away from the detector center (within a distance of 0.5~m from  the inner surface of the acrylic vessel).
For further reduction of the accidental background formed by the coincidence of two uncorrelated events, prompt-delayed pairs are required to occur within a 1.0~ms coincidence window and within a spatial distance of 1.5~m.
The selection criteria will be further tuned when JUNO begins data taking and the actual accidental background is measured.

A muon veto strategy has been developed to suppress backgrounds induced by neutrons and long-lived isotopes, such as $^9$Li/$^8$He produced by cosmic muons.
This strategy, as reported in Ref.~\cite{JUNO:2022mxj}, uses a variable muon veto time window depending on the proximity of the candidate event to a recent muon track or spallation neutron capture.
The details are as follows:
\begin{itemize}
    \item
    A veto time window of 1~ms after a muon passes through the water pool Cherenkov detector and/or the CD is applied over the entire fiducial volume to suppress spallation neutrons and short-lived radioisotopes.
    This veto time window is effective against spallation neutrons with an average capture time of 200~\textmu s.

    \item
    For well-reconstructed muon tracks in the CD caused by a single muon or two far-apart muons (\textgreater 3~m), a veto time window of 0.6, 0.4, and 0.1~s is applied to the candidate events with reconstructed vertices closer than 1, 2, and 4~m away from each of the muon track, respectively.

    \item For events containing two close and parallel muons (\textless 3~m), which constitute approximately 0.6\% of muon-related events, a single track is often reconstructed. 
    A veto is applied around this track as described earlier, but the cylinder's radii increase according to their separation, which can be inferred from the charge pattern around the entrance and exit points on the CD.
   
    \item
    For events in which a track cannot be properly reconstructed, which amount to about 2\% of all muon-related events and occur primarily when more than two muons go through the detector simultaneously, a 0.5~s veto time window is applied over the entire fiducial volume.

    \item 
    A 1.2~s veto is applied to all candidate events reconstructed inside a 3~m radius sphere around spallation neutron capture events. 
    This cut aids in further rejecting backgrounds from cosmogenic isotope decays topically produced with accompanying spallation neutrons.
\end{itemize}

The estimated IBD detection efficiency after all selection cuts is 82.2\%,
which is independent of the antineutrino energy. 
A breakdown of the cut efficiencies and the corresponding IBD rates are shown in Table~\ref{tab:efficiency}, which is the same as the result in Ref.~\cite{JUNO:2022mxj}.
The uncertainties of the IBD selection efficiency are studied based on the expected performance of the reconstruction as evaluated with the simulation.
The uncertainty owing to the fiducial volume cut is 0.4\%, estimated by considering a 2~cm vertex reconstruction bias and the correction in the radial direction.
The uncertainty of the selection cuts, including the prompt energy, delayed energy, coincidence time, and distance cuts, is estimated to be 0.2\%.
The muon veto has negligible uncertainty on the IBD rate because the rejected time window and detector volume can be calculated precisely.
The number of target protons in the detector determines the IBD reaction rate and has an uncertainty of 0.9\%, estimated from the Daya Bay experiment~\cite{An:2016srz}, which uses a similar LS.
These uncertainties constitute a $1\%$ detector normalization uncertainty.
This absolute IBD detection efficiency uncertainty has a negligible effect on the NMO sensitivity, which relies entirely on the spectral shape information.

\begin{table}[!htb]
    \centering
    \begin{tabular}{lccc}
    \hline
    Selection Criterion & Efficiency [\%] & IBD Rate [day$^{-1}$]\\
    \hline
    All IBDs						& 100.0 	& 57.4 \\
    Fiducial Volume 						& 91.5 	& 52.5 \\
    IBD Selection 				& 98.1		& 51.5 \\
    ~~~~Energy Range 					& ~~~~99.8		& - \\
    ~~~~Time Correlation ($\Delta T_{p-d}$) 		& ~~~~99.0 	& - \\
    ~~~~Spatial Correlation ($\Delta R_{p-d}$) 	& ~~~~99.2		& - \\
    Muon Veto (Temporal$\oplus$Spatial)		& 91.6 	& 47.1  \\
    \hline
    Combined Selection					& 82.2 	&  47.1\\
    \hline
    \end{tabular}
    \caption{Summary of cumulative reactor antineutrino selection efficiencies. The reported IBD rates (with baselines \textless 300~km) refer to the expected events per day after the selection criteria are progressively applied. These rates are calculated for the nominal reactor power and baselines listed in Table~\ref{tab:Reactor}.
    }
    \label{tab:efficiency}
\end{table}

\subsection{Residual Backgrounds}
\label{sec:background}
 
After the application of the IBD selection criteria, about 10\% of the IBD candidates are background events that are caused by five main sources:

\begin{itemize}
    \item Radiogenic events, such as $\alpha$, $\beta$, and $\gamma$ decays from natural radioactivity in the detector material and in the adjacent environment.
    \item Cosmogenic events, such as fast neutrons and unstable isotopes produced by the interactions of cosmic muons with detector and surrounding materials, typically via spallation.
    \item Atmospheric neutrinos, i.e., neutrinos of all flavors created during the collisions of primary cosmic rays with the Earth's atmosphere.
    \item Electron antineutrinos emitted by distant (\textgreater 300~km) reactors.
    \item Electron antineutrinos created in the uranium and thorium decay chains in the Earth, i.e., geoneutrinos.
\end{itemize}

These five sources produce the seven categories of backgrounds listed in Table~\ref{tab:background} and the estimated spectra are presented in Figure~\ref{fig:JUNO:signal_and_background}.
The accidental background is formed by the coincidence of two uncorrelated events.
The prompt-like signals primarily consist of radiogenic events, whereas the delayed-like signals primarily consist of radiogenic events and spallation neutrons produced by cosmic muons.
The rate and spectrum of the accidental background are estimated using the energy, position, and rate of uncorrelated events from radioactivity, as presented in Ref.~\cite{JUNO:2021kxb}, which uses the latest information on the detector materials.
After all the selection criteria, the remaining accidental background rate is estimated to be 0.8/day, with a 1\% rate uncertainty and negligible shape uncertainty owing to the precise energy spectrum measurement using an off-window method.

Muons passing through the detector can produce the long-lived cosmogenic isotopes $^9$Li and $^8$He, whose correlated $\beta$--$n$ decays mimic the IBD signature.
The production yields of $^9$Li and $^8$He are estimated to be 127/day and 40/day, respectively, based on the JUNO detector simulation and measurements from the KamLAND and Borexino experiments~\cite{Abe:2009aa, Bellini:2013pxa, Bellini:2008mr}.
The production of these elements is strongly correlated in time and space with the parent muon. 
After the muon veto strategy described in Section~\ref{sec:selection} is applied, the residual $^9$Li/$^8$He background rate is estimated to be 0.8/day with an uncertainty of 20\%.
The spectral shape of the $^9$Li/$^8$He background is evaluated based on a calculation benchmarked at Daya Bay~\cite{DayaBay:2016ggj} with a 10\% uncertainty.

The fast neutron background is produced by energetic neutrons generated by cosmic muons.
The neutrons scatter off protons and are then captured, thereby mimicking the IBD prompt and delayed signals, respectively.
Owing to the relatively short neutron capture time, the muon veto time window can easily reject most of the fast neutrons, if muons are tagged by the muon veto system.
However, some fast neutrons cannot be vetoed if their parent muons only pass through rocks around the detector.
The rate of the residual fast neutron background is estimated to be 0.1/day based on a simulation, with a 100\% uncertainty.
The fast neutron spectrum is assumed to be flat in the energy range of interest with a 20\% uncertainty. 
These assumptions are reasonable as observed in both simulation and recent reactor experiments~\cite{DoubleChooz:2019qbj, RENO:2018dro, DayaBay:2016ggj}.

The ${}^{13}$C($\alpha$,n)${}^{16}$O background is formed by the interactions of $\alpha$ particles from natural radioactivity with the $^{13}$C in the LS.
The radiation emitted by the ${}^{16}$O excitation state mimics the IBD prompt signal, whereas neutron capture emulates the delayed signal.
The background rate is estimated to be 0.05/day with a 50\% uncertainty based on the studies in Ref.~\cite{An:2015jdp}.
The spectral shape is obtained from a simulation~\cite{DayaBay:2016ggj} with a 50\% uncertainty assumed to be the same as the rate uncertainty.
A new estimation of the ${}^{13}$C($\alpha$,n)${}^{16}$O background~\cite{Mendoza:2019vgf} has been performed with the latest JUNO software and will be published after this paper. 
The impact on the NMO sensitivity has been checked and is found to be negligible because the ${}^{13}$C($\alpha$,n)${}^{16}$O background rate is much smaller than other backgrounds.

Geoneutrinos are electron antineutrinos produced by radioactive decay chains of U and Th inside the Earth. 
The total rate of the geoneutrinos is estimated to be 1.2/day in the JUNO detector with a 30\% rate uncertainty and 5\% spectral shape uncertainty.
These values of the rate and uncertainties are consistent with the results of our previous work~\cite{An:2015jdp}. 
The rate is adjusted by $+0.1$~day$^{-1}$ owing to the increase of the muon veto efficiency in this study.

World reactors whose distances are larger than 300~km are quoted as background because they do not contribute to the NMO sensitivity. 
The energy spectrum and event rate are estimated according to the world nuclear reactor information provided in Ref.~\cite{international2019iaea} with a contribution of 1.0/day.
We set the rate and spectral uncertainties to be 2\% and 5\%, respectively.

Atmospheric neutrinos interact with the nuclei in the JUNO detector via either charged or neutral current interactions. 
Refs.~\cite{Cheng:2020aaw,Cheng:2024uyj} indicate that the final states of neutral current interactions may induce correlated backgrounds.
The energy spectrum and event rate depend strongly on the interaction generators used in the estimation.
The background rate is estimated to be 0.16/day, and the spectral shape is evaluated using GENIE~2.12.0~\cite{Andreopoulos:2009rq} as the nominal model.
Both the rate and shape uncertainties are assigned to be 50\% to cover the variation in results from the different generators.

The residual background rates and their uncertainties are summarized in Table~\ref{tab:background}. 
They are the same as those in Ref.~\cite{JUNO:2022mxj}.
In the absence of well-motivated models that can predict the correlation between bins in these empirical estimates, all background shape uncertainties in this study are treated as bin-to-bin uncorrelated, which enables the spectra to vary in any possible configuration within the specified uncertainty envelopes. 

\begin{table}[tbh!]
    \centering
    \begin{tabular}{
    l
    S[table-format=1.2]
    S[table-format=1.2]
    S[table-format=3]
    S[table-format=2]
    }
    \hline
    Backgrounds & 
    \multicolumn{1}{c}{Rate [day$^{-1}$]} & 
    \multicolumn{1}{c}{B/S [\%]} & 
    \multicolumn{1}{c}{Rate Unc. [\%]} & 
    \multicolumn{1}{c}{Shape Unc. [\%]} \\
    \hline
    Geoneutrinos 		& 1.2 & 2.5 & 30 & 5 \\
    World reactors 		& 1.0 & 2.1 & 2 & 5 \\
    Accidentals 		& 0.8 & 1.7 & 1 & \multicolumn{1}{c}{negligible} \\
    $^9$Li/$^8$He 	    & 0.8 & 1.7 & 20 & 10 \\
    Atmospheric neutrinos 		& 0.16 & 0.34 & 50 & 50 \\
    Fast neutrons 		& 0.1 & 0.21 & 100 & 20 \\
    ${}^{13}$C($\alpha$,n)${}^{16}$O & 0.05 & 0.01 & 50 & 50 \\
    Total backgrounds  & 4.11 & 8.7 & &  \\
    \hline
    \end{tabular}%
    \caption{Expected background rates, background to signal ratio (B/S), and rate and shape uncertainties. 
    The B/S ratio is calculated using the IBD signal rate of 47.1/day from Table~\ref{tab:Reactor}.
    }
    \label{tab:background}%
\end{table}%

\section{TAO Experimental Setup} 
\label{sec:tao}

TAO (also known as JUNO-TAO) is a JUNO satellite experiment with an energy resolution better than 2\% at 1~MeV~\cite{JUNO:2020ijm}.
Its major goal is to provide a reference spectrum for JUNO and eliminate any possible model dependence in the determination of the NMO.
TAO is located approximately 44~m from one reactor core (TS-C1) of the Taishan NPP and approximately 217~m from the other reactor core (TS-C2).
These two Taishan reactor cores contribute more than 99.99\% of TAO's antineutrino signals, with TS-C2 contributing 4\%.
TAO consists of a central detector, an outer shielding, and a veto system.
The central detector contains 2.8 t of gadolinium-doped LS filled in a spherical acrylic vessel with an inner radius of 0.9 m as the target volume, with the same hydrogen atom mass fraction and abundance as JUNO.
Charged particles in TAO are detected through scintillation and Cherenkov processes, which generate photons detected by 4024 silicon photomultipliers (SiPMs) covering approximately 10~m$^2$  with $\sim$50\% photon detection efficiency.
The detector operates at $-50$~$^\circ$C to reduce the SiPM dark noise to an acceptable level (less than 100~Hz/mm$^2$).
The central detector is surrounded by 1.2~m thick water tanks on the sides, 1~m thick high density polyethylene (HDPE) on the top, and 10~cm thick lead at the bottom to shield against ambient radioactivity and cosmogenic neutrons. 
Cosmic muons are detected by the veto system including a water tank instrumented with PMTs and a plastic scintillator array on the top.
The TAO detector will begin operating around the same time as the JUNO detector.
For more information about the TAO detector, please refer to Ref.~\cite{JUNO:2020ijm}.

TAO's detector response and expected IBD signal and backgrounds, along with the uncertainties, are discussed in the following sections.

\subsection{TAO Detector Response}
\label{sec:tao:detresp}

The TAO detector response model is constructed similar to that of JUNO, as described in Section~\ref{sec:response}. 
In addition to the energy transfer, nonlinearity, and resolution effects, the energy leakage effect, which is the loss of energy due to prompt events depositing part of their energy in non-active volumes, is considered specifically for TAO owing to its much smaller size with respect to JUNO. 
A spherical fiducial volume cut of 25~cm from the LS boundary is applied to reduce energy leakage and mitigate backgrounds. 
The modelling of the energy leakage, nonlinearity, and resolution effects are introduced in detail in the following.

The energy leakage effect is encoded in a dedicated matrix, where the columns and rows represent the true positron energy (including the annihilation gammas) and deposited energy in the active volume of the detector, respectively.
The two-dimensional matrix is constructed by running the TAO detector simulation with fixed-energy positrons from 0.8~MeV to 12~MeV uniformly distributed in the detector and extracting the deposited energy distribution.
This results in a 2240$\times$2240 energy leakage matrix with an energy bin width of 5~keV, wihch is sufficiently reasonable compared with the 20~keV bin width adopted in the final fitting.

The LSNL effect is modeled similar to that of JUNO, as described in Section~\ref{subsec:lsnl}, and its uncertainty is treated as correlated with JUNO.
This is based on the fact that the recipes of the JUNO and TAO LSs are almost identical, as well as the nonlinearity calibration schemes~\cite{JUNO:2020xtj,Xu:2022mdi}, in which similar calibration sources and enclosures are used.
Furthermore, no change on the nonlinearity curve is found for the TAO LS at $-50\,^{\circ}\mathrm{C}$ based on a comparison of the measured light yield curves of the Compton electrons induced by the 0.511~MeV and 1.275~MeV gammas at $20\,^{\circ}\mathrm{C}$, $-20\,^{\circ}\mathrm{C}$, and $-50\,^{\circ}\mathrm{C}$ with a $^{22}$Na source.
An additional nuisance parameter is introduced to parameterize a possible difference in energy scales between the two detectors.
This uncorrelated uncertainty takes a conservative value of 0.5\%, which is
selected based on the fact that both JUNO and TAO can determine the energy peaks of neutron capture on hydrogen with a precision of $0.1-0.2$\% by using neutron calibration sources~\cite{JUNO:2020xtj,Xu:2022mdi}. 
The energy response non-uniformity of TAO introduces a 0.3\% uncertainty on the energy scale.

TAO’s energy resolution is the same as that in Ref.~\cite{JUNO:2020ijm} but with two additional effects included: the energy response non-uniformity effect and the fluctuation in Cherenkov light emission~\cite{Xu:2022mdi}. 
Using a three-dimensional calibration of the detector response~\cite{Xu:2022mdi}, the non-uniformity can be corrected with a residual value of 0.2\%, which has a small impact on the energy resolution.
The Cherenkov light emission simulation is implemented in the TAO detector using the same method as in the JUNO detector~\cite{JUNO:EnergyResolutionPaper}. 
Despite these improvements, the new resolution curve shown in the inset of Figure~\ref{fig:reponese:pdfs} is very similar to that of Ref.~\cite{JUNO:2020ijm}.

\begin{figure}[tbh!]
    \centering
    \includegraphics[width=0.9\textwidth]{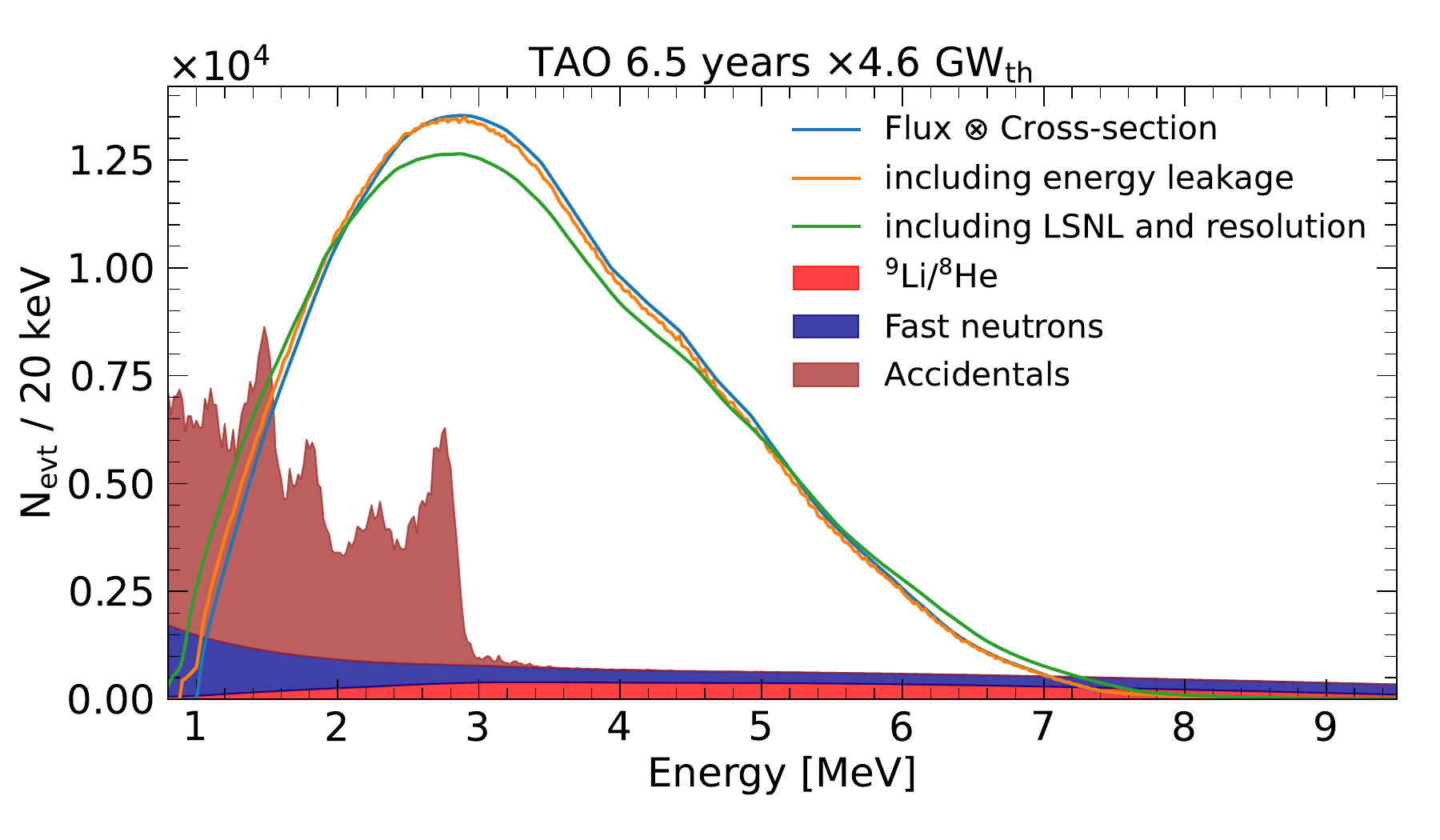}
    \caption{
    Energy spectra of expected IBD signals in TAO from both TS-C1 and TS-C2 with energy leakage, LSNL, and energy resolution effects applied. The reconstructed energy spectra expected for the three major backgrounds are represented by the filled histograms. All spectra corresponds to an exposure of 6.5 years $\times$ 4.6~GW$_{\rm th}$.
    The curve including energy leakage has small wiggles owing to the fluctuation in the the energy leakage in the simulation. 
    However, the wiggles are invisible after the energy resolution is applied.
     }
    \label{fig:TAO:signal_and_background}
\end{figure}

The predicted energy spectra after incorporating different effects in the energy response of the TAO detector are shown in Figure~\ref{fig:TAO:signal_and_background}.
The IBD positron energy spectrum is calculated with the reactor antineutrino flux from both TS-C1 and TS-C2, convoluted with the IBD cross section.
Subsequently, the energy leakage effect is applied, which causes an energy shift owing to energy loss, followed by the the effects of LSNL and energy resolution in sequence.

\subsection{TAO Signal and Backgrounds}

The reactor antineutrino flux emitted from the Taishan reactor cores is observed simultaneously by TAO and JUNO. 
In the prediction of the IBD signal, TAO differs from JUNO in terms of the baseline, target mass, live-time, and detection efficiency, while sharing the same IBD cross section and reactor antineutrino flux model, as described in Section~\ref{sec:flux}. The averaged fission fractions of the antineutrino flux for both detectors are also very similar.

TAO has a baseline of approximately 44~m from the core of TS-C1 as its main source, which means that basically no oscillation effect will occur in the standard three-flavor neutrino oscillation model. 
The target mass of TAO is 2.8 t with a fiducial mass of approximately 1 t.
The detection efficiency, which includes the IBD selection efficiency and fiducial volume cut, is calculated from the TAO detector simulation to be 17\%. 
The veto system results in a dead time of 9.6\%, as evaluated from the simulation with the measured cosmic muon rate.
In total, TAO will observe approximately 1000 neutrino signals per day in the fiducial volume. 
A conservative relative rate uncertainty of 10\% is assigned to the detection efficiency from the fiducial cut uncertainty, which has a negligible impact on the result of the JUNO-TAO combined fit, which is introduced in Section~\ref{sec:analysis_results}.
The reactor-related uncertainties of the Taishan cores, such as reactor power and fission fraction uncertainties, are fully correlated with JUNO.

The signal spectrum is derived from the Daya Bay-corrected Huber-Mueller model and with TAO’s detector response, as described in Section~\ref{sec:tao:detresp}. 
The signal spectra, with and without detector effects, are shown in Figure~\ref{fig:TAO:signal_and_background}.
In the JUNO-TAO combined fit, each bin of the initial reactor antineutrino spectrum is kept unconstrained to eliminate reactor flux model dependence and is fully correlated between JUNO and TAO.
Owing to the energy leakage effect, the signal shape differs according to choices of fiducial volume cut. 
Accordingly, the uncertainty in vertex reconstruction is propagated to the signal shape uncertainty, which is estimated to be less than 0.5\% in most of the energy regions.
An additional uncertainty uncorrelated for each bin is assigned to consider the fact that JUNO observes a group of reactors of different types with asynchronized burn-up evolution, whereas TAO observes primarily one Taishan core.
This uncertainty is estimated to be 0.35\% based on the available fission fraction data from Taishan NPP and the antineutrino spectral shape difference between $^{235}$U and $^{239}$Pu.

Three main backgrounds are considered for TAO: fast neutron and $^9$Li/$^8$He backgrounds induced by cosmic muons and accidental background primarily caused by the random coincidence of natural radioactivity events with single neutron captures. The $^9$Li/$^8$He background rate has been determined through theoretical calculation, whereas the other two background rates are estimated from the TAO detector simulation~\cite{Li:2022wqc}. All the background rates are summarized in Table~\ref{tab:taosigbkg}.
The fast neutron background spectrum is taken from the TAO detector simulation, whereas the spectra of the other two backgrounds are assumed to mimic those of JUNO. This approach is suitable for this sensitivity study as they will be measured using data-driven methods.
The energy spectra expected for the three major backgrounds are shown in Figure~\ref{fig:TAO:signal_and_background}.
Given the higher IBD rate due to the shorter baseline and smaller detector size, the other backgrounds relevant to JUNO become negligible for TAO.

For the $^9$Li/$^8$He background, a rate uncertainty of 20\% and shape uncertainty of 10\% uncorrelated between each bin are assumed, similar to JUNO.
The rate and spectrum of the accidental and fast neutron backgrounds can be evaluated with data-driven methods.
An average of 1 month of reactor-off data per year is expected to provide a direct measurement of these backgrounds.
The reactor-off data can constrain the accidental background well; thus, only a rate uncertainty of 1\% is assigned. 
With more than 3 years of data taking, the reactor-off data will provide sufficient fast neutron statistics at a rate of 2000 events per day, when the muon veto selection is not applied.
The statistical uncertainty of this control sample is assigned to each bin of the fast neutron background spectrum, which is expected to be the dominant uncertainty and generally less than 10\%. The neutrino signal contamination from TS-C2 to this control sample is percent-level; thus, it is neglected.

Table~\ref{tab:taosigbkg} summarizes the signal and background rates and shape models for TAO, as well as the uncertainties considered.

\begin{table}[ht!]
\begin{center}
\begin{tabular}{lS[table-format=3]S[table-format=2]cc}
    \hline
    Type   & \multicolumn{1}{c}{Rate [day$^{-1}$]} & \multicolumn{1}{c}{Rate Uncert. [\%]} & Shape Model & Shape Uncert. [\%] \\
    \hline
    Signal & 1000 & 10 & same as JUNO & FF, FV, ES \\
    Fast neutron & 86 & \multicolumn{1}{c}{--} & TAO simulation & \textless10\% \\
    $^9$Li/$^8$He & 54 & 20 & same as JUNO & \phantom{\textless}10\% \\
    Accidental & 190 & 1 & same as JUNO & --  \\
    \hline
\end{tabular}
\caption {TAO signal and background rates, as well as shapes and their uncertainties. 
For the signal shape uncertainty, only the effects that have not been considered for the JUNO signal are explicitly indicated: FF refers to the additional fission fraction uncertainty introduced to TAO to cover the difference in burn-up history observed by TAO and JUNO, FV refers to the uncertainty due to vertex reconstruction uncertainty introduced by the fiducial volume cut, and ES refers to the relative energy scale uncertainty to consider potential differences between the calibration efforts of TAO and JUNO.
}
	\label{tab:taosigbkg}%
\end{center}
\end{table}%

\section{Analysis and Results}
\label{sec:analysis_results}

\subsection{Inputs and models}

\onlyinsubfile{\section{Analysis}\subsection*{None (introduction)}}

The analysis is performed by three separate groups. 
Each analysis group has implemented an independent software for producing the predictions for the JUNO and TAO detectors, managing the systematic uncertainties, and performing the statistical analysis. 
The groups performed out detailed cross-checks at all stages of the analysis, from the prediction of the energy spectra to the determination of the sensitivity to the oscillation parameters and the NMO.
Consistent results were obtained, and only one set of results is presented in the paper.

The groups use the same inputs, which are further referred to as the common inputs. 
These include the antineutrino spectra from the reactors, IBD cross section, detector efficiency, energy response, expected backgrounds, and other parameters characterizing the performance of the nuclear reactors and antineutrino detectors. 
Compared with Ref.~\cite{JUNO:2022mxj}, several updates that became available after the publication have been incorporated.
Most notably, the full chain of the JUNO offline software to perform detector simulation, electronics simulation, waveform reconstruction~\cite{Lin:2022htc}, and event reconstruction~\cite{Li:2021oos,Huang:2021baf} is employed to predict the LSNL and energy resolution.
We include updates to the detector geometries, photon detection efficiency (PDE) from PMT mass testing data~\cite{JUNO:2022hlz}, scintillation quenching effect, Cherenkov light yield, PMT optical model~\cite{Wang:2022tij}, and LS optical model.
These updates have a negligible impact on the precision of the oscillation parameters in Ref.~\cite{JUNO:2022mxj} but are essential to the NMO sensitivity by improving the energy resolution from 3\% to 2.95\% at 1 MeV.
More details about the evaluation of the new energy nonlinearity and resolution are available in Ref.~\cite{JUNO:EnergyResolutionPaper}.
The implementation of the systematic uncertainties in energy nonlinearity and resolution in this study follows a strategy similar to that in Ref.~\cite{JUNO:2022mxj}.

\subfileref


\onlyinsubfile{\section{Analysis}\subsection{Model}}

The prediction of the expected spectrum follows the methods described in \cref{sec:response,sec:tao}.  
The result of the prediction represents the number of events in a histogram binned into 340 bins with 20~keV intervals in the majority of energy range, as shown in \cref{tab:final_binning}. 
Wider bins are used such that each of them can provide at least 500 events to ensure that $\chi^2$ is unbiased.
\NewText{\begin{table}[tbh]
    \centering
    \begin{tabular}{lcccccccccccccccccc}
       &  &  &  &  &  &  &  &  &  &  &  &  &  &  &  &  &  &   \\
       \toprule
         Bin edges [MeV]
         & \multicolumn{2}{c}{0.8}
         & \multicolumn{2}{c}{0.94}
         & \multicolumn{2}{c}{\dots}
         & \multicolumn{2}{c}{7.44}
         & \multicolumn{2}{c}{\dots}
         & \multicolumn{2}{c}{7.8}
         & \multicolumn{2}{c}{\dots}
         & \multicolumn{2}{c}{8.2}
         & \multicolumn{2}{c}{12} \\
        Bin width [keV]  &
                   & \multicolumn{2}{|c}{140}
                   & \multicolumn{4}{|c}{20}
                   & \multicolumn{4}{|c}{40}
                   & \multicolumn{4}{|c}{100}
                   & \multicolumn{2}{|c|}{3800}
                   \\
        Number of bins  &
                   & \multicolumn{2}{|c}{1}
                   & \multicolumn{4}{|c}{325}
                   & \multicolumn{4}{|c}{9}
                   & \multicolumn{4}{|c}{4}
                   & \multicolumn{2}{|c|}{1} \\
       \bottomrule
    \end{tabular}
    \caption{Definition of the bins of the histograms for the IBD prompt energy used in the analysis for both JUNO and TAO. The total number of bins is 340 for each detector.}
    \label{tab:final_binning}
\end{table}}

Although the groups are using similar inputs and methods, the methods of integrating the IBD kinematics are different; they are explained in more detail in the Appendix.
However, the three groups for both the JUNO and TAO predictions agree closely. 
In terms of the spectra, the relative difference is $\lesssim$$\num{e-5}$ in 90\% of the energy range, excluding the low and high energies, where the numbers of events are low. In terms of the rate, the relative difference is also $\lesssim$$\num{e-5}$. 

\subfileref

\subsection{Statistical Methods}
\label{subsec:stat}

\onlyinsubfile{\section{Analysis}\subsection{Statistical methods}}

The aim of this study is to estimate the median sensitivity of the JUNO experiment to the neutrino mass ordering.
The statistical methods in a frequentist approach are thoroughly discussed  in Refs.~\cite{Qian:2012zn,Blennow:2013oma}.
In general, the test statistic is defined as $\DTestStat=\Min\TestStat[IO](D) - \Min\TestStat[NO](D)$, where $D$ is the data and \TestStat is the minimized function, which is 
minimized over all the parameters, assuming two mass orderings.
The critical value \TestStatC is selected such that $\DTestStat\geq\TestStatC$ corresponds to the normal mass ordering; otherwise, it corresponds to the inverted mass ordering.
The sensitivity or confidence level of the test is defined as ($1-\alpha$), where $\alpha=\alpha(\TestStatC)$ is the probability of rejecting the null hypothesis $H_0$ (e.g. normal mass ordering) if $H_0$ is true. 
We use the notion of ``median sensitivity'', which is defined under the assumption that, given the same value of \TestStatC, the probability $\beta(\TestStatC)$ to accept the null hypothesis $H_0$ equals $50\%$ if the alternative hypothesis $H_1$ (e.g., inverted mass ordering) is true.
The probability is expressed in terms of the number of standard deviations \SensN,
such that $\alpha(n)=\frac{2}{\sqrt{2\pi}}\int_n^\infty e^{-x^2/2}dx$.

In this study, the sensitivity is estimated in a Gaussian limit, where
the median sensitivity is defined as $\SensN=\sqrt{\left|\DTestStatA\right|}$, with \DTestStatA being the value of the test statistic for expectation without fluctuations (Asimov data).
On average, $\DTestStat>0\ (<0)$ when the data follows normal (inverted) mass ordering. 
For the Asimov data \DataA, $\Min\TestStat[NO](\DataA[NO])=\Min\TestStat[IO](\DataA[IO])=0$.

The groups use three different definitions of \TestStat, either based on the $\chi^2$ function or approximately equal to it. 
While the details are covered in the Appendix, we consider the construction based on \DChi in the following text, which is defined as
\begin{multline}
  \TestStat{}(\ParsFree, \SinSqTheta13, \ParsNuisance, \ParsNuisanceCorr)=
  \sum_\IdxDet
    \left(\Theory\TDet(\ParsFree, \SinSqTheta13, \ParsNuisance, \ParsNuisanceCorr)-\Datai\TDet\right)^T
    \left(V\TDet\tstat + V\TDet\tbtob\right)^{-1}
    \left(\Theory\TDet-\Datai\TDet\right)
  + \\
  + \ChiSqOsc(\SinSqTheta13)
  + \ChiSqNuisance(\ParsNuisance)
  + \ChiSqNuisanceCorr(\ParsNuisanceCorr)
  ,
  \label{eq:chi2-a-b}
\end{multline}
where $\ParsFree$ is a set of free parameters, including \Dm31, \Dm32, and \SinSqTheta12; $\ParsNuisance$ is a set
of all the uncorrelated nuisance parameters, listed in \cref{tab:input_unc,tab:input_unc_juno,tab:input_unc_tao}; and $\ParsNuisanceCorr$ is a set of partially correlated parameters (\cref{tab:input_unc}). 
The index $d$ enumerates the JUNO and TAO detectors. 
Column $\Theory$ contains the expected event spectrum, and column $\Data$ contains the data.
The diagonal matrix $V\tstat$ contains the statistical uncertainties, defined according to the Combined Neyman-Pearson (CNP) \cite{Ji:2019yca} approach:
$\left(V\tstat\right)_{ii} = 3/\left({1/\Theory_i}+{2/\Datai_i}\right)$. 
The matrix $V\tbtob$ contains bin-to-bin uncertainties:
$V\tbtob^\text{JUNO}$ is diagonal and contains background shape uncertainties of the JUNO detector, whereas
$V\tbtob^\text{TAO}$ contains background shape uncertainties of the TAO detector, uncertainties due to the fiducial volume
cut, and extra fission fraction uncertainties. 
Most of the listed bin-to-bin uncertainties are uncorrelated between the bins and thus contribute only to the diagonal. 
The uncertainties owing to the the fiducial volume cut are partially correlated between the bins and thus make $V\tbtob^\text{TAO}$ non-diagonal.

The nuisance part is defined as $\ChiSqNuisance(\ParsNuisance)=\sum_i (\ParNuisance_i - \overline{\ParNuisance_i})^2 / \sigma^2(\ParNuisance_i)$
for the uncorrelated parameters and as
$\ChiSqNuisanceCorr(\ParsNuisanceCorr)=(\ParNuisanceCorr - \overline{\ParNuisanceCorr})^T V^{-1}_{\ParNuisanceCorr} (\ParNuisanceCorr - \overline{\ParNuisanceCorr})$
for the correlated ones (fission fractions), with $\ParsNuisanceCorr$ used as a column.
Here, $\overline{\ParNuisance_i}$ and $\overline{\ParNuisanceCorr}$ denote central values,
$\sigma(\ParNuisance_i)$ is the uncertainty of the $i$-th nuisance parameter, and $V_\ParNuisanceCorr$ is
the covariance matrix of a set of correlated parameters. The parameters themselves are discussed in the following section.
$\sin^2\theta_{13}$ is constrained in the term $\ChiSqOsc(\SinSqTheta13)$ with central value and uncertainty obtained from PDG2020~\cite{Zyla:2020zbs:PDG2020Release}$: \sin^2\theta_{13}=(2.18\pm 0.07)\times 10^{-2}$.

The parameters of the reactor antineutrino spectrum are contained either in $\ParsFree$, where they are treated without constraints (totally free), or in $\ParsNuisance$, when they are treated with deliberately large uncertainties (almost free).

The models of the JUNO and TAO experiments, containing the common reactor antineutrino part, are analyzed
in a simultaneous fit that acts as a far/near measurement: the TAO data are more sensitive to the reactor antineutrino spectrum, whereas the JUNO data are sensitive to neutrino oscillations.
An alternative mode exists, where only the JUNO detector is considered, and the shape uncertainty of the antineutrino spectrum due to the TAO measurement is implemented as a relative uncertainty of each bin of the IBD histogram after the detector effects are applied. 
The corresponding nuisance parameters are added to $\ChiSqNuisance$.
The alternative approach follows the analysis procedure defined in Ref.~\cite{JUNO:2022mxj} and is used as a cross-check, producing consistent results.


A dedicated Monte-Carlo study is performed based on the complete model by each group to test the applicability of the Gaussian limit
for two cases: a) no systematic uncertainties and b) full treatment of the systematic uncertainties.
100,000 pseudoexperiments generated for each NMO revealed that $\SensN=\sqrt{\left|\DTestStatA\right|}$
provides a consistent estimation of the median sensitivity. As \TestStat approximately follows the $\chi^2$ function, we will further refer to the test statistics as \DChi, and we obtain $\Delta \chi^2_\text{min} = \DTestStatA$ for Asimov data after minimizing over all of the parameters in the test statistic.

Although the statistical methods are primarily developed for estimating the NMO sensitivity, they can also be utilized to estimate uncertainties on the oscillation parameters that agree with the results in Ref.~\cite{JUNO:2022mxj}.

\subfileref

\subsection{Systematic Uncertainties}

\onlyinsubfile{\section{Analysis}\subsection{Uncertainties}}

The free parameters of the analysis include \Dm31, \Dm21, and \SinSqTheta12, while $\SinSqTheta13$ is constrained as mentioned in the previous section.

The parameters describing the shape of the energy spectrum of reactor antineutrinos are common for JUNO and TAO.
As the fit of the common spectrum is an important part of the analysis, multiple options are used.
In one case, the correction to the antineutrino spectrum is applied with 400 bins with a width of 20~keV starting from 1.8~MeV and a 100\% uncertainty is applied via the nuisance terms. 
In another case, the correction is applied with a piece-wise exponential, defined on 149 segments of varying width starting from 1.8~MeV. 
The widths of the segments are selected to be as wide as 1.5 of the width of the energy resolution of the TAO detector at corresponding \Evis. 
A set of 150 parameters that control the values at the edges of the segments are fit without constraints applied. Other options were also used during the analysis, all yielding consistent results.

\begin{table}[htb!]
\newcommand\CellCombination{\multicolumn{1}{c}{\{\dots\}\hspace*{2ex}}}
\centering
\begin{tabular}{lll<{\hspace*{-1cm}}rSSS}
  \toprule
  \multicolumn{3}{l}{Systematic effect} &
  \multicolumn{1}{p{20mm}}{\raggedright Number of parameters} &
  \multicolumn{3}{c}{Relative uncertainty [\%]} \\
  \cmidrule{5-7}
  {}                                    & {}                                          &                              &                  & \multicolumn{1}{c}{Input}            & \multicolumn{1}{c}{JUNO rate} & \multicolumn{1}{c}{TAO rate}                 \\
  \midrule
  \multicolumn{3}{l}{Total uncertainty}                                                                              & \{73+680\}       &                                      & 2.5                           & 10                                           \\
  {}                                    & \multicolumn{2}{l}{Statistical uncertainty}                                &                  &                                      & 0.31                          & 0.078                                        \\
  {}                                    & \multicolumn{2}{l}{Systematic uncertainty}                                 & \{73+680\}       &                                      & 2.5                           & 10                                           \\
  \multicolumn{3}{l}{Common}                                                                                         & \{56\}           &                                      & 2.0                           & 2.2                                          \\
  {}                                    & \multicolumn{2}{l}{Reactors}                                               & \{52\}           &                                      & 2.0                           & 2.2                                          \\
  {}                                    & {}                                          & Baselines                    & $-$\phantom{)}   & 0                                    & 0                             & 0                                            \\
  {}                                    & {}                                          & Reactor flux normalization   & 1\phantom{)}     & 2.0                                  & 2.0                           & 2.0                                          \\
  {}                                    & {}                                          & Thermal Power                & 9\phantom{)}     & 0.5                                  & 0.17                          & 0.48                                         \\
  {}                                    & {}                                          & Spent Nuclear Fuel           & 1\phantom{)}     & 30                                   & 0.084                         & 0.085                                        \\
  {}                                    & {}                                          & Non-equilibrium neutrinos    & 1\phantom{)}     & 30                                   & 0.19                          & 0.19                                         \\
  {}                                    & {}                                          & Mean energy per fission      & 4\phantom{)}     & \multicolumn{1}{c}{$0.13-0.25$}      & 0.090                         & 0.090                                        \\
  {}                                    & {}                                          & Fission fractions (part. correlated)  & 36\phantom{)}    & 5.0               & 0.25                          & 0.63                                         \\
  {}                                    & \multicolumn{2}{l}{LSNL}                                                   & 4\phantom{)}     & 100\phantom{)}                       & 0.0081                        & 0.0028                                       \\
  \multicolumn{3}{l}{Individual}                                                                                     &                  &                                      & 1.4                           & 10                                           \\
  \bottomrule
\end{tabular}
\caption{
  Summary of the systematic effects relevant for both JUNO and TAO detectors.
  The input uncertainties represent the constraints applied to the relevant nuisance parameters or the bin-to-bin uncertainties of the relevant spectrum.
  For JUNO, rate uncertainties are given relative to the expected IBD rate from the Taishan, Yangjiang, and Daya Bay reactors, whereas for TAO, uncertainties are given relative to the expected IBD rate from the Taishan reactors for the DAQ time required to reach 3$\sigma$ sensitivity to the NMO.
  The individual rate uncertainties for JUNO and TAO are 1.4\% and 10\%, respectively.
  Curly braces \{\} indicate that the line provides a summary of a group of uncertainties, which are detailed below with extra indentation.
  The number of parameters in a group is a sum of the numbers of parameters, corresponding to the next indent level.
  In total, the number of uncorrelated parameters is 73, and the number of partially correlated parameters is 680.
  The uncertainties of the fission fractions are partially correlated between four isotopes in each reactor core.
  Some of the systematic effects are propagated via bin-to-bin uncertainties, which are combined for each bin, and thus forming 340 bins in total.
}
\label{tab:input_unc}
\end{table}

\begin{table}[htbp]
\newcommand\CellCombination{\multicolumn{1}{c}{\{\dots\}\hspace*{2ex}}}
\centering
\begin{tabular}{lllrSS}
  \toprule
  \multicolumn{3}{l}{Systematic effect} &
  \multicolumn{1}{p{20mm}}{\raggedright Number of parameters} &
  \multicolumn{2}{c}{Relative uncertainty [\%]} \\
  \cmidrule{5-6}
  {}                                    & {}                                          &                                              &                  & \multicolumn{1}{c}{Input}                      & \multicolumn{1}{c}{Rate}  \\
  \midrule
  \multicolumn{3}{l}{JUNO systematic uncertainty}                                                                                    & \{13+340\}       &                                                & 1.4                          \\
  {}                                                                                  & \multicolumn{2}{l}{\SinSqTheta13}            & 1\phantom{)}     & 3.2\phantom{)}                                 & 0.13                         \\
  {}                                                                                  & \multicolumn{2}{l}{Matter density (MSW)}     & 1\phantom{)}     & 6.1\phantom{)}                                 & 0.062                        \\
  {}                                    & \multicolumn{2}{l}{Detector normalization}                                                 & 1\phantom{)}     & 1.0                                            & 1.0                          \\
  {}                                    & \multicolumn{2}{l}{Energy resolution}                                                      & 3\phantom{)}     & \multicolumn{1}{c}{0.77, 1.6, 3.3}   & 5.4e-07                      \\
  {}                                                                                  & \multicolumn{2}{l}{Background rates}         & \{7\}            &                                                & 0.96                         \\
  {}                                    & {}                                          & Accidentals                                  & 1\phantom{)}     & 1.0                                            & 0.019                        \\
  {}                                    & {}                                          & \nuclide[9]{Li}/\nuclide[8]{He}              & 1\phantom{)}     & 20                                             & 0.37                         \\
  {}                                    & {}                                          & Fast neutrons                                & 1\phantom{)}     & 100                                            & 0.23                         \\
  {}                                    & {}                                          & \nuclide[13]{C}$(\alpha, n)$\nuclide[16]{O}  & 1\phantom{)}     & 50                                             & 0.058                        \\
  {}                                    & {}                                          & Geoneutrinos                                 & 1\phantom{)}     & 30                                             & 0.84                         \\
  {}                                    & {}                                          & Atmospheric neutrinos                        & 1\phantom{)}     & 50                                             & 0.19                         \\
  {}                                    & {}                                          & World reactors                               & 1\phantom{)}     & 2.0                                            & 0.046                        \\
  {}                                                                                  & \multicolumn{2}{l}{Background shape}         & 340\phantom{)}   &                                                & 0.033                        \\
  {}                                    & {}                                          & \nuclide[9]{Li}/\nuclide[8]{He}              & (340)            & 10                                             & 0.012                        \\
  {}                                    & {}                                          & Fast neutrons                                & (340)            & 20                                             & 0.0026                       \\
  {}                                    & {}                                          & \nuclide[13]{C}$(\alpha, n)$\nuclide[16]{O}  & (340)            & 50                                             & 0.0053                       \\
  {}                                    & {}                                          & Geoneutrinos                                 & (340)            & 5.0                                            & 0.026                        \\
  {}                                    & {}                                          & Atmospheric neutrinos                        & (340)            & 50                                             & 0.011                        \\
  {}                                    & {}                                          & World reactors                               & (340)            & 5.0                                            & 0.010                        \\
  \midrule
  \multicolumn{3}{m{0.3\linewidth}}{IBD spectrum shape\hspace{\fill}\newline \mbox{uncertainty} from TAO}                            & 340\phantom{)}   & \multicolumn{1}{c}{$1.3-45$}                   & 0.35                         \\
  \bottomrule
\end{tabular}
\caption{
  Summary of the systematic effects impacting only the JUNO detector.
  The input uncertainties represent the constraints applied to the relevant nuisance parameters or the bin-to-bin uncertainties of the relevant spectrum.
  The rate uncertainties are given relative to the expected IBD rate from the Taishan, Yangjiang, and Daya Bay reactors for the DAQ time required to reach 3$\sigma$ sensitivity to the NMO.
  Curly braces \{\} indicate that the line provides a summary of a group of uncertainties, which are detailed below with extra indentation.
  The number of parameters in a group is a sum of the numbers of parameters, corresponding to the next indent level.
  The number of uncorrelated parameters is 13, and the number of partially correlated parameters is 340, which corresponds to the number of bins.
  Some of the systematic effects are propagated via bin-to-bin uncertainties, which are combined for each bin, thus forming 340 bins in total.
  Round brackets () are used to indicate the individual sources of bin-to-bin uncertainties.
  The IBD spectrum shape uncertainty from TAO is not used in the main analysis; therefore, its impact on the rate is not included in the total uncertainty.
}
\label{tab:input_unc_juno}
\end{table}

\begin{table}[htbp]
\newcommand\CellCombination{\multicolumn{1}{c}{\{\dots\}\hspace*{2ex}}}
\centering
\begin{tabular}{lllrSS}
  \toprule
  \multicolumn{3}{l}{Systematic effect} &
  \multicolumn{1}{p{20mm}}{\raggedright Number of parameters} &
  \multicolumn{2}{c}{Relative uncertainty [\%]} \\
  \cmidrule{5-6}
  {}                                    & {}                                          &                                       &                  & \multicolumn{1}{c}{Input}            & \multicolumn{1}{c}{Rate}  \\
  \midrule
  \multicolumn{3}{l}{TAO}                                                                                                     & \{4+340\}        &                                      & 10                          \\
  {}                                    & \multicolumn{2}{l}{Detector normalization}                                          & 1\phantom{)}     & 10                                   & 10                          \\
  {}                                    & \multicolumn{2}{l}{Energy scale}                                                    & 1\phantom{)}     & 0.50                                 & 0.0018                      \\
  {}                                    & \multicolumn{2}{l}{Accidentals' rate}                                               & 1\phantom{)}     & 1.0                                  & 0.20                        \\
  {}                                    & \multicolumn{2}{l}{\nuclide[9]{Li}/\nuclide[8]{He} rate}                            & 1\phantom{)}     & 20                                   & 1.2                         \\
  {}                                    & \multicolumn{2}{l}{Bin-to-bin}                                                      & 340\phantom{)}   &                                      & 0.086                       \\
  {}                                    & {}                                          & \nuclide[9]{Li}/\nuclide[8]{He} shape & (340)            & 10                                   & 0.038                       \\
  {}                                    & {}                                          & Fast neutrons                         & (340)            & \multicolumn{1}{c}{$0.36-5.0$}       & 0.016                       \\
  {}                                    & {}                                          & Fission fractions difference          & (340)            & 0.35                                 & 0.023                       \\
  {}                                    & {}                                          & Fiducial volume                       & (340)            & \multicolumn{1}{c}{$0.091-14$}       & 0.072                       \\
  \bottomrule
\end{tabular}
\caption{
  Summary of the systematic effects impacting only the TAO detector.
  The input uncertainties represent the constraints applied to the relevant nuisance parameters or the bin-to-bin uncertainties of the relevant spectrum.
  The rate uncertainties are given relative to the expected IBD from the Taishan reactors for the DAQ time, required to reach 3$\sigma$ sensitivity to the NMO.
  Curly braces \{\} indicate that the line provides a summary of a group of uncertainties, which are detailed below with extra indentation.
  The number of parameters in a group is the sum of the numbers of parameters corresponding to the next indent level.
  Some of the systematic effects are propagated via bin-to-bin uncertainties, which are combined for each bin, thus forming 340 bins in total.
  Round brackets () are used to indicate the individual sources of bin-to-bin uncertainties.
  The uncertainties due to the fiducial volume cut are partially correlated between the bins.
}
\label{tab:input_unc_tao}
\end{table}

A summary of the systematic effects relevant for both JUNO and TAO detectors is given in \cref{tab:input_unc}. 
The breakdown of the effects, exclusive to the JUNO and TAO detectors is presented in \cref{tab:input_unc_juno,tab:input_unc_tao}, respectively. 
All the uncertainties listed in these three tables are considered in the combined analysis.
In the second column of the tables, the number of uncorrelated or partially correlated parameters is reported. The input uncertainty, shown in the third column,
corresponds either to the uncertainty applied for the input parameter via the nuisance term or the relevant spectrum when added to the bin-to-bin covariance matrix.
The value from each systematic group is propagated to the final event rate and shown in the last and second-last columns (\cref{tab:input_unc}) as a ratio to the IBD rate.
The bin-to-bin (shape) uncertainties are defined as corrections to the heights of corresponding histograms.
For example, the fast neutron relative shape uncertainty $\delta_\text{f.n.}=20\%$ is applied to the fast neutron spectrum $H_i^\text{f.n.}$ to build the diagonal covariance matrix $V_{ii}^\text{f.n.}=(\delta_\text{f.n.}H_i^\text{f.n.})^2$; the bin-to-bin uncertainty due to the difference of the fission fractions is applied to the IBD bins, etc.

\begin{figure}[htbp]
  \centering
    \includegraphics[width=0.9\linewidth]{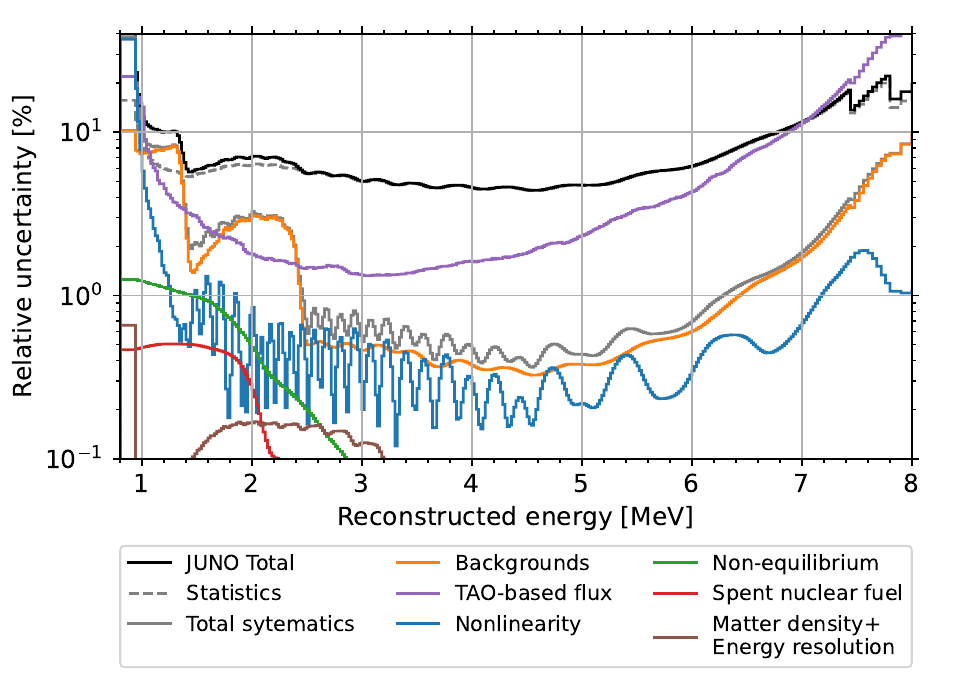}
  \caption{
Shape uncertainties of the predicted spectrum of the JUNO detector, presented relative to the number of IBD events from Taishan, Yangjiang, and Daya~Bay reactors in each bin for the DAQ time required to reach 3$\sigma$ sensitivity to NMO. The absolute uncertainties are obtained by generating simulated samples, where systematic parameters are varied based on their assumed uncertainties and taking square roots of diagonal elements of the resulting covariance matrices. The rate uncertainties of the spent nuclear fuel and non-equilibrium corrections, as well as of the backgrounds, also distort the observed spectrum and are consequently included in this figure. The square of the total uncertainty is the quadratic sum of all the individual uncertainties, except the TAO based flux uncertainty, which is not used in the main analysis.
}
  \label{fig:shape:diagonal_term_juno}
\end{figure}

\begin{figure}[htbp]
  \centering
    \includegraphics[width=0.9\linewidth]{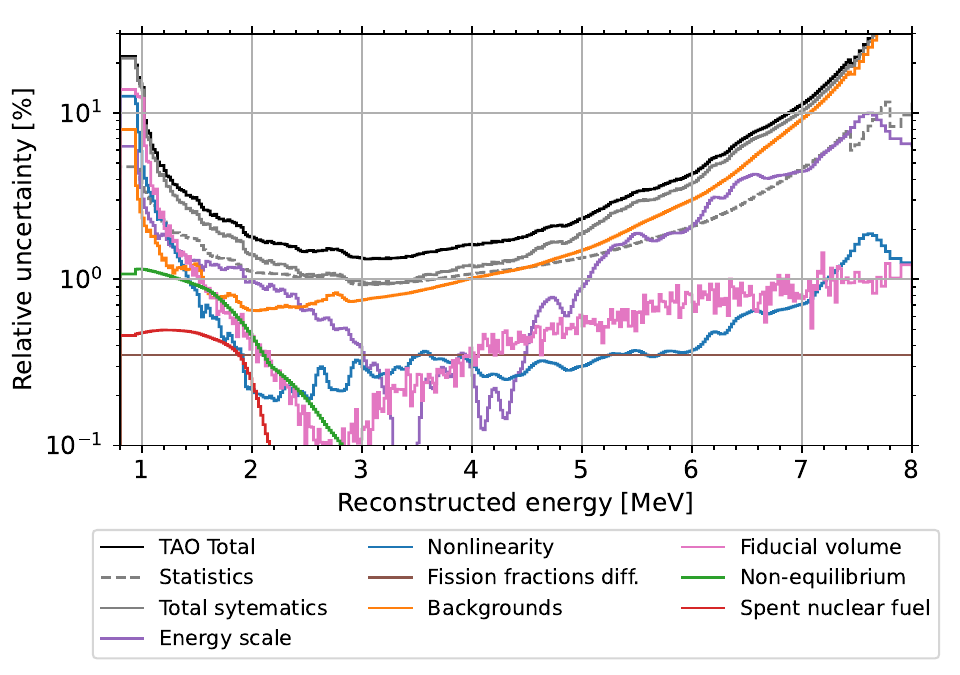}
  \caption{
Shape uncertainties of the predicted spectrum of the TAO detector, presented relative to the number of IBD events from Taishan reactors in each bin for the DAQ time required to reach 3$\sigma$ sensitivity to NMO. The absolute uncertainties are obtained by generating simulated samples, where systematic parameters are varied based on their assumed uncertainties and taking square roots of diagonal elements of the resulting covariance matrices. The rate uncertainties of the spent nuclear fuel and non-equilibrium corrections, as well as of the backgrounds, also distort the observed spectrum and are consequently included in this figure. The square of the total uncertainty is a quadratic sum of all the individual uncertainties.
  }
  \label{fig:shape:diagonal_term_tao}
\end{figure}

The spectrum shape uncertainties are shown in \cref{fig:shape:diagonal_term_juno,fig:shape:diagonal_term_tao} for the JUNO and TAO detectors, respectively. 
These uncertainties are obtained by propagating the input uncertainties to the final spectrum, computing the covariance matrix via differentiation, and obtaining square roots of its diagonal terms.
These uncertainties are reported for illustration purposes only as they are not used in the analysis directly.
The diagonal part of the nonlinearity covariance matrix has visible oscillations, which are caused by the interplay between neutrino oscillations, variation in the spectrum due to nonlinearity, and choice of the bin sizes. 
For TAO, the features due to the interplay of nonlinearity and bin sizes are less pronounced in the absence of oscillations. 
The impact of the energy scale on the TAO spectrum has a minimum at approximately 3.4~MeV, as it is expected to be close to the mode of the histogram. 
Its position is slightly offset to the higher energies owing to the asymmetry of the histogram. 
The minimum of the uncertainty due to the fiducial volume cut has similar origins as the energy scale shift. 
The uncertainty due to the fiducial volume cut is computed via MC and thus contains fluctuations. The fluctuations are then propagated to the IBD spectrum and affect the energy scale uncertainty.

\label{fig:taospec}
\subfileref

\FloatBarrier 

\subsection{Results}

\onlyinsubfile{\section{Analysis}\subsection{Results}}

With the nominal configuration, statistical methods, and uncertainty sources described in this work, we numerically calculate JUNO's expected NMO sensitivity for the Asimov dataset assuming JUNO and TAO start data taking simultaneously.
Because the three analysis groups produce consistent $|\Delta\chi^2_\text{min}|$ estimations within a relative 0.5\% difference, only one result is shown here.

\begin{figure}[hbt]
  \centering
  \includegraphics[width=0.9\textwidth]{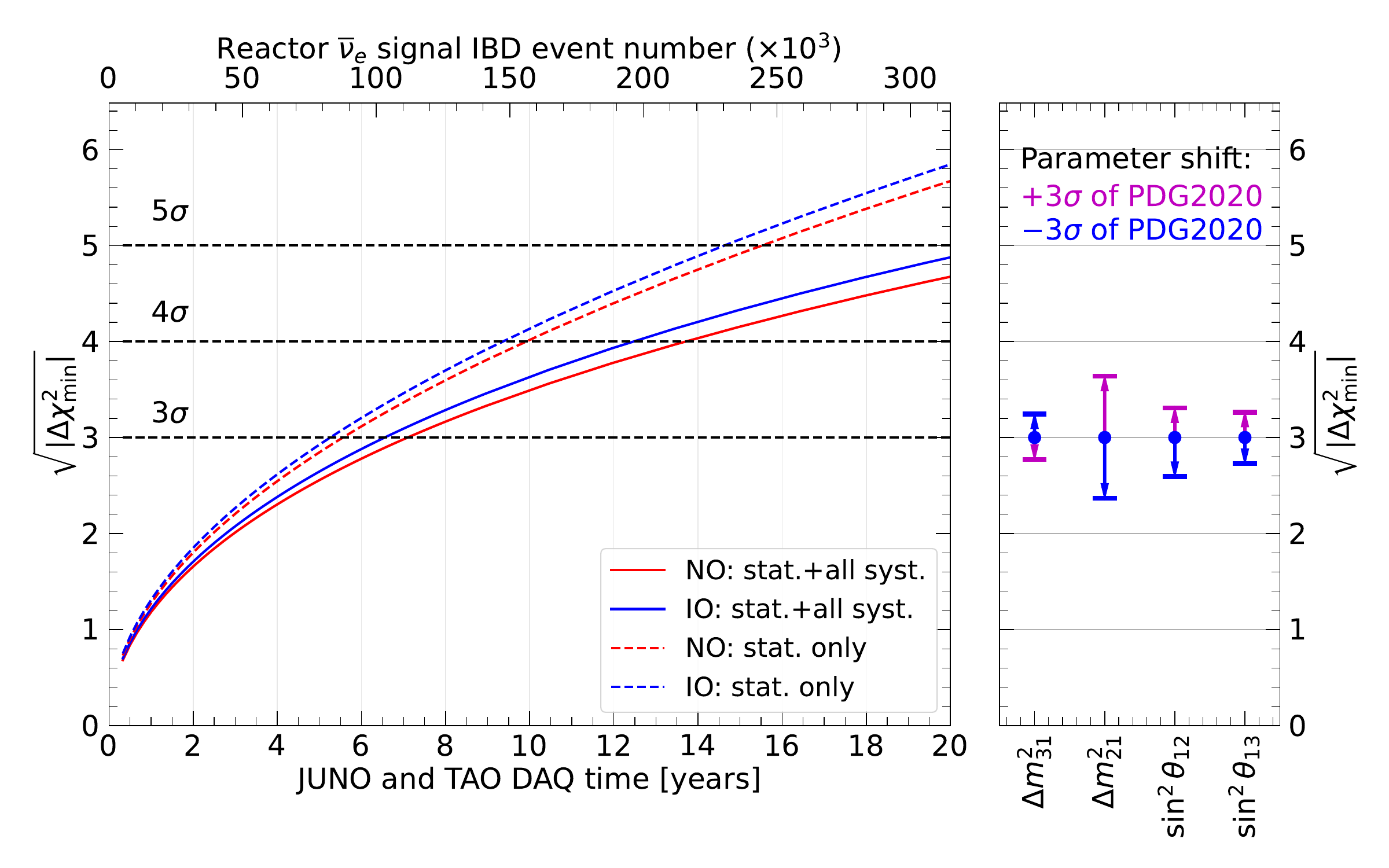}
  \caption{ NMO discriminator $\Delta\chi^2_\text{min}$ as a function of JUNO and TAO data-taking times for both NO (red) and IO (blue). 
  The horizontal black dashed lines represent $3\sigma$, $4\sigma$, and $5\sigma$ significances. 
  The solid lines are for the cases of full systematic uncertainties, and the dashed lines are for the statistical-only case. The 11/12 reactor duty cycle is considered in the conversion of exposure to the data-taking time. 
  We can observe that, after 7.1~years of data taking, JUNO can determine the neutrino mass ordering with $3\sigma$ significance when NO is true. If IO is true, it is 3.1$\sigma$ under the same exposure. 
  We assume that JUNO and TAO begin data taking at the same time. 
  The right panel shows the sensitivity dependence on the true values of the oscillation parameters, evaluated by shifting the values 3$\sigma$ (of PDG2020~\cite{Zyla:2020zbs:PDG2020Release}) from the nominal values. The results are presented for the normal ordering for the exposure needed by JUNO to reach $3\sigma$ sensitivity.
}
  \label{fig:sensitivityT:JUNOTAO}
\end{figure}

$|\Delta\chi^2_\text{min}|$ is slightly less than 9 for 6 years of reactor antineutrino data at JUNO for both the normal and inverted mass orderings. 
After 7.1~years of data taking with an assumed 11/12 duty factor for the reactors (an exposure of 6.5~years $\times$ 26.6~GW$_{\rm th}$), JUNO has a median NMO sensitivity of $3\sigma$ ($3.1\sigma$) for the normal (inverted) mass ordering.
Figure~\ref{fig:sensitivityT:JUNOTAO} shows the median NMO sensitivity as a function of JUNO and TAO data-taking times for both NO and IO hypotheses and for the cases with and without all systematic uncertainties.
We find that the sensitivity is primarily driven by statistical uncertainty, resulting in $\Delta \chi^2_\text{min}$ approximately following a linear function of exposure. 
This relationship enables converting variations in $\Delta \chi^2_\text{min}$ into the corresponding adjustments of data-taking time required to reach a median NMO sensitivity of $3\sigma$.

The analysis is performed with the Asimov data produced with the oscillation parameters from PDG~2020~\cite{Zyla:2020zbs:PDG2020Release}. 
Note that the previous published sensitivity~\cite{An:2015jdp} of JUNO was based on PDG~2014~\cite{Agashe:2014kda:PDG2014}.
To demonstrate the effect of the variation in the oscillation parameters on the NMO sensitivity, another study is performed by generating Asimov data with the oscillation parameters shifted from the nominal values described in Section~\ref{sec:intro}.
We scan the oscillation parameters within $\pm 15\%$ to cover the 3-sigma region around their central values.
We find that $|\Delta\chi^2_\text{min}|$ is positively correlated with $\sin^2\theta_{12}$, $\sin^2\theta_{13}$, and $\Delta m^2_{21}$ and anti-correlated with $\Delta m^2_{31}$. 
The right panel of Figure~\ref{fig:sensitivityT:JUNOTAO} shows the impact of the oscillation parameter true values on the NMO sensitivity.

\begin{figure}[tb!]
    \centering
    \includegraphics[width=0.8\textwidth]{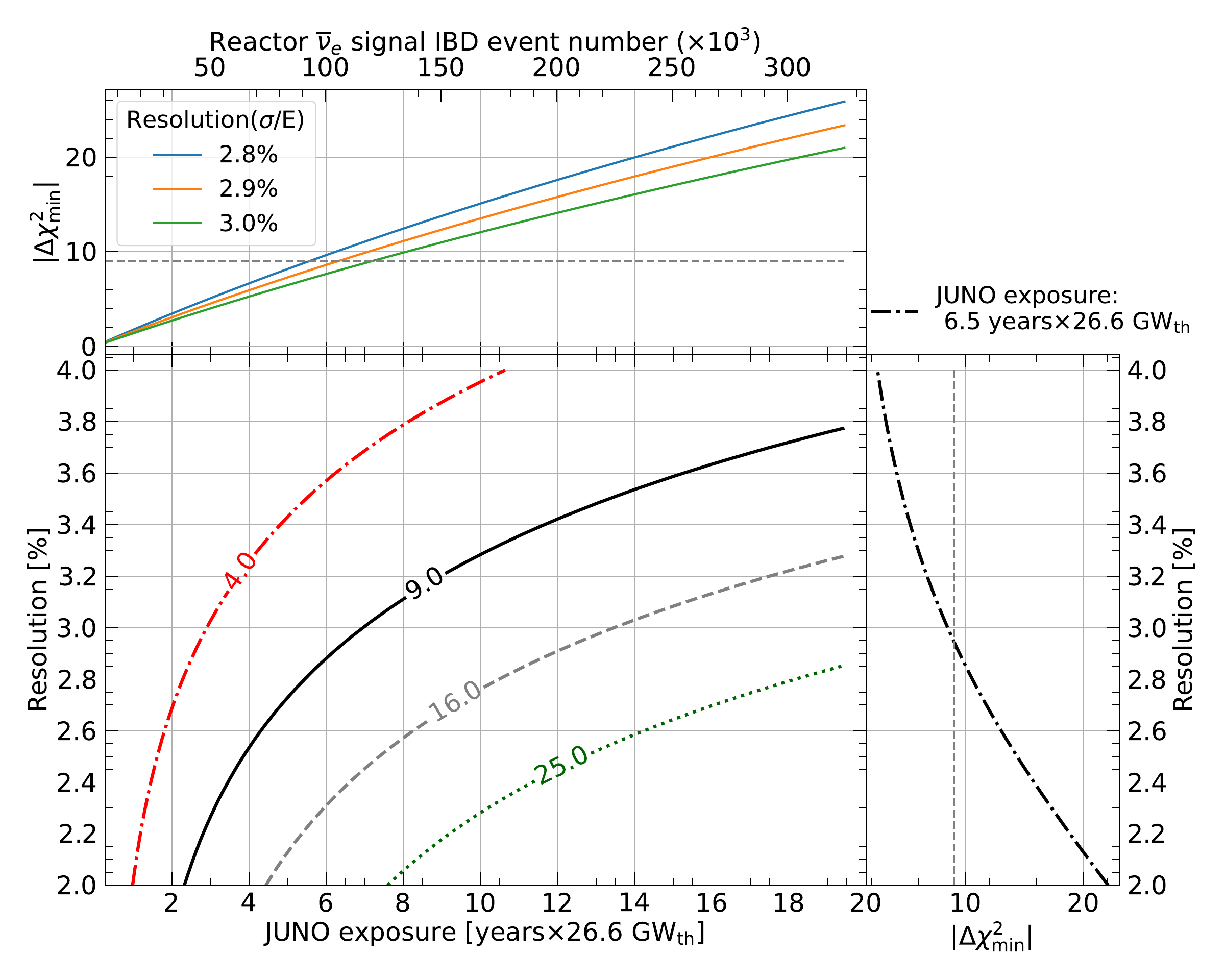}
    \caption{Contours of $|\Delta\chi^2_\text{min}|$ as a function of exposure and energy resolution at 1~MeV under the assumption of NO. The resolution is scanned by varying $a$ and fixing $b=0.64\times10^{-2}$, $c=1.20\times10^{-2}$~MeV.  
  The black, gray, and green contour lines denote $3\sigma$, $4\sigma$, and $5\sigma$ significance levels, respectively. The top panel shows the time evolution of the $|\Delta\chi^2_\text{min}|$ for the energy resolution of $2.8\%$, $2.9\%$, and $3.0\%$ at 1~MeV. The right panel shows the required energy resolution to achieve different $|\Delta\chi^2_\text{min}|$ under the JUNO exposure of 6.5~years$\times 26.6\ {\rm GW_{th}}$ (data-taking time of 7.1~years with a reactor duty cycle of 11/12). }
    \label{fig:sensitvity:time-resolution-dependece}
\end{figure}

The energy resolution and exposure are crucial factors for determining the NMO. We present the $|\Delta\chi^2_\text{min}|$ contours as a function of exposure and energy resolution in Figure~\ref{fig:sensitvity:time-resolution-dependece}. The top panel shows the dependence on exposure for energy resolutions of 2.8\%, 2.9\%, and 3.0\% at 1~MeV. 
The right panel shows $|\Delta\chi^2_\text{min}|$ as a function of energy resolution, with the nominal exposure of JUNO reaching a $3\sigma$ significance. 
Although we have a more realistic detector response model, the dependence is similar to our previous study~\cite{An:2015jdp}.

Table~\ref{tab:sensitivity:breakdown} shows the breakdown of the impact of the systematic uncertainties on the NMO sensitivity for 7.1 years of data taking.
The sensitivity is determined by cumulatively considering the statistical uncertainty of the reactor antineutrino sample and each source of systematic uncertainty.
Only the rows with backgrounds and the total sensitivity consider the statistical uncertainty of the backgrounds, whereas the other rows consider only the statistical uncertainty of the IBD sample.
We can observe that the dominant systematics are the backgrounds, reference spectrum uncertainty, and nonlinearity uncertainty. 
If no external constraint is used for \SinSqTheta13, $\Delta\chi^2_\text{min}$ decreases by another 0.4 (0.2) for the case of NO (IO).

\begin{table}[htb!]
    \centering
    \begin{tabular}{l@{~}lS[table-format=2.2]S[table-format=+1.2]}
        \toprule
        \multicolumn{2}{l}{Uncertainties}                            & \multicolumn{1}{c}{$|\Delta\chi^2_\text{min}|$}         & \multicolumn{1}{c}{$|\Delta\chi^2_\text{min}|$ change} \\
        \midrule
        \multicolumn{2}{l}{Statistics of JUNO and TAO}                               & 11.5                             &  \\
                                     & + Common uncertainty     & 10.8                             & -0.7  \\
                                             & + TAO uncertainty     & 10.2                             & -0.6  \\
                                          & + JUNO geoneutrinos           & 9.7                              & -0.5  \\
                                          & + JUNO world reactors         & 9.4                              & -0.3  \\
                                          & + JUNO accidental             & 9.2                              & -0.2  \\
                                          & + JUNO $^9$Li/$^8$He          & 9.1                              & -0.1  \\
                                          & + JUNO other backgrounds      & 9.0                              & -0.05 \\
        \multicolumn{2}{l}{Total}                                    & 9.0                              &  \\
        \bottomrule
    \end{tabular}%
    \caption{Relative impact of individual sources of uncertainty on the NMO sensitivity. 
    The sensitivity is determined by cumulatively considering the statistic uncertainty of the reactor antineutrino sample and each source of systematic uncertainty.  
    The common uncertainty includes the systematic uncertainties listed in Table~\ref{tab:input_unc}. The TAO uncertainty includes the systematic uncertainties shown in Table~\ref{tab:input_unc_tao}. 
    The ``$|\Delta\chi^2_\text{min}|$ change'' column represents the decrease in $|\Delta\chi^2_\text{min}|$ compared with the value in previous row. 
    The other backgrounds include the atmospheric neutrino background, fast neutron background, and ${}^{13}$C($\alpha$,n)${}^{16}$O background. 
    The total sensitivity resulting from simultaneously considering all sources of error is shown in the last row of the table. 
    All uncertainties correspond to an exposure to 6.5 years $\times$ 26.6~GW thermal power for JUNO.}
    \label{tab:sensitivity:breakdown}%
\end{table}%

\begin{figure}[tb!]
    \centering
        \includegraphics[width=0.8\textwidth]{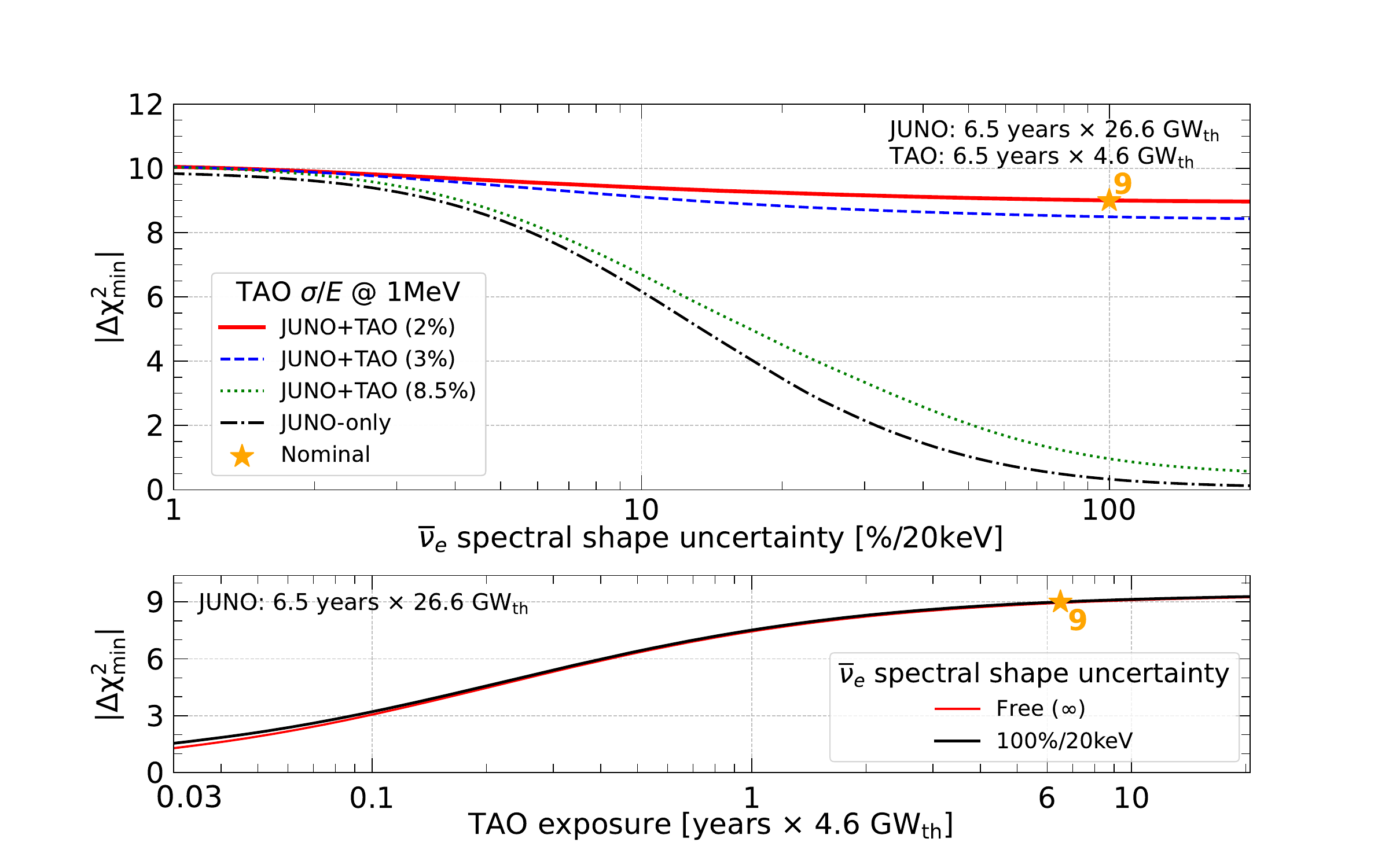}
    \caption{ NMO sensitivity examination under various configurations of antineutrino spectral shape uncertainty, TAO energy resolution, and exposure. Top: Nominal JUNO+TAO analysis results (marked as a star) use an exposure of 6.5~years$\times 26.6\ {\rm GW_{th}}$, nominal TAO energy resolution ($\sim$2\% at 1 MeV), and 100\% antineutrino spectral shape uncertainty. The NMO sensitivity (red) is scanned as functions of antineutrino spectral shape uncertainty. 
    For cases of poor energy resolution, the NMO sensitivity is also calculated by changing the TAO energy resolution to the values of Daya Bay ($\sim$8.5\% at 1 MeV) and JUNO ($\sim$3\% at 1 MeV). 
    Without the TAO detector, the sensitivity of the JUNO-only case (black dot-dashed line) is also studied for comparison. Bottom: JUNO+TAO sensitivity versus TAO exposure time given a fixed exposure for JUNO in the cases of 100\% antineutrino spectral shape uncertainty (black) and free spectral shape (red). }
    \label{fig:sensitvity:tao-importance}
\end{figure}

Figure~\ref{fig:sensitvity:tao-importance} depicts the importance of the TAO detector in NMO sensitivity by considering alternative configurations of the antineutrino spectral shape uncertainty, energy resolution, and exposure time.
In a JUNO-only analysis without TAO, the NMO sensitivity diminishes as the antineutrino spectral shape uncertainty increases, emphasizing the dependence of the sensitivity on the antineutrino flux model.
In contrast, the model-independent analysis of JUNO+TAO maintains the sensitivity even with antineutrino spectral shape uncertainty of larger than 100\%, as long as TAO achieves a similar (3\%) or better (2\%) energy resolution than JUNO. 
However, a TAO energy resolution at the level of Daya Bay detector (8.5\%) results in almost total loss of the NMO sensitivity in the absence of external constraints on the antineutrino spectral shape.
TAO exposure is another important factor, and the bottom panel of Figure~\ref{fig:sensitvity:tao-importance} shows that a running time of few years is required to maximize the NMO sensitivity.

\section{Conclusion}
\label{sec:conclusion}
JUNO and TAO detectors are in an advanced construction stage and are expected to begin commissioning in near future.
The JUNO detector, located 52.5~km from the Taishan and Yangjiang reactors, will collect approximately 16 thousand IBD signals per year with a 20~kton LS target.
The TAO detector will be placed 44~m from one of the Taishan reactor cores, and it will measure the reactor antineutrino energy spectrum with negligible neutrino oscillation.
Their unprecedented energy resolutions will enable it to make a precise measurement of the antineutrino energy spectrum for identifying of the NMO.

In this study, the NMO sensitivity is assessed using the most recent information available about the location and overburden of the experimental sites, local and global nuclear reactors, responses of both detectors, signal and background estimations, and systematic uncertainties.
Three independent analyses are conducted, all beginning with the same common inputs and producing consistent results on sensitivity.
The median sensitivity to reject the wrong mass ordering is found to be 3$\sigma$ with an exposure to 6.5 years $\times$ 26.6~GW thermal power, assuming the normal ordering is true. 
The corresponding data-taking time is 7.1 years, under the assumption of an 11/12 duty factor for the reactors.
If the inverted ordering is true, the median sensitivity is $3.1\sigma$ for the same exposure. 

This study focuses only on JUNO's sensitivity to the NMO using reactor antineutrinos. 
However, it demonstrated that incorporating external constraints on $\Delta m_{32}^2$ from other experiments, as well as information from atmospheric neutrinos in JUNO, could enhance the NMO sensitivity~\cite{Li:2013zyd, An:2015jdp, Nunokawa:2005nx}.
JUNO will play a unique role in identifying the NMO by employing the LS detector technology, without relying on matter effects or having any dependency on $\delta_\text{\tiny CP}$ and $\theta_{23}$.
Thus, these measurements will be unique and have a high degree of complementarity to those pursued by other experiments, increasing confidence in NMO determination and enabling effective tests on the three-neutrino framework.

\section*{Acknowledgement}

We are grateful for the ongoing cooperation from the China General Nuclear Power Group.
This work was supported by
the Chinese Academy of Sciences,
the National Key R\&D Program of China,
the CAS Center for Excellence in Particle Physics,
Wuyi University,
and the Tsung-Dao Lee Institute of Shanghai Jiao Tong University in China,
the Institut National de Physique Nucl\'eaire et de Physique de Particules (IN2P3) in France,
the Istituto Nazionale di Fisica Nucleare (INFN) in Italy,
the Italian-Chinese collaborative research program MAECI-NSFC,
the Fond de la Recherche Scientifique (F.R.S-FNRS) and FWO under the ``Excellence of Science – EOS” in Belgium,
the Conselho Nacional de Desenvolvimento Cient\'ifico e Tecnol\`ogico in Brazil,
the Agencia Nacional de Investigacion y Desarrollo and ANID - Millennium Science Initiative Program - ICN2019\_044 in Chile
the Charles University Research Centre and the Ministry of Education, Youth, and Sports in Czech Republic,
the Deutsche Forschungsgemeinschaft (DFG), the Helmholtz Association, and the Cluster of Excellence PRISMA+ in Germany,
the Joint Institute of Nuclear Research (JINR) and Lomonosov Moscow State University in Russia,
the joint Russian Science Foundation (RSF) and National Natural Science Foundation of China (NSFC) research program,
the MOST and MOE in Taiwan,
the Chulalongkorn University and Suranaree University of Technology in Thailand,
University of California at Irvine and the National Science Foundation in the US.

\cleardoublepage

\appendix
\section*{Appendix: differences between the analyses}
\label{sec:diff_appendix}
\addcontentsline{toc}{section}{Appendix: differences between the analyses}

\onlyinsubfile{\section{Analysis}\subsection{Model}}

The prediction of the expected spectrum follows the methods described in \cref{sec:response,sec:tao}. 
The process of prediction proceeds from neutrino energy \Enu to reconstructed energy \Erec via a chain:\\
\hspace*{\fill}%
\begin{tikzpicture}[
    baseline=-0.7mm,
    sub/.style={font=\relsize{-2},midway,below=3mm,text width=3cm,align=center}
    ]
  \matrix (en) [matrix of math nodes,nodes={},column sep=7mm]
  {
  \Enu & \Epos (\Enu, \cos\theta) & \Edep (\Epos) 
  & \Evis (\Edep) 
  & \Erec (\Evis)
  ,
  \\
  };
  \foreach \label/\idx\idxnext in {kinematics/1/2,{annihilation+leakage}/2/3,LSNL/3/4,resolution/4/5}
    \draw[-stealth] (en-1-\idx) -- (en-1-\idxnext) node [sub] {\label};
\end{tikzpicture}
\hfill\refstepcounter{equation}(\theequation) \label{eq:energy_conversion} \\
which, while closely related to the processes inside the detector, is approximated in a treatment of the detector effects, most notably of the gammas from the annihilation.

More than 98\% of the neutrino energy \Enu is converted to the positron energy \Epos in the IBD process~\cite{Vogel:1999zy,Strumia:2003zx}. 
After annihilation with an electron, the total deposited energy \Edep{} consists of the kinetic energy of the positron and two electron masses, which are contributed by the two gammas from the annihilation. 
At this stage, the differences in the detector response to the positron and gammas are ignored. 
For the TAO prediction, the energy leakage is considered during this step. 
Thereafter, by applying the distortion due to LSNL, different degrees of quenching for the positrons and gammas~\cite{Adey:2019zfo:DYB_NL} are considered. 
Finally, a finite energy resolution is considered to provide the reconstructed energy \Erec.
The result represents the number of events in a histogram binned into 340 bins with 20~keV intervals in most of the energy range, as shown in \cref{tab:final_binning}. 
Wider bins are used so each of them can provide at least 500 events, ensuring that $\chi^2$ is unbiased.

The core difference between the groups, denoted as A, B, and C, lies in their way of handling the kinematics of the IBD interaction.
The generalized formulae for the groups are shown in \crefrange{eq:int_a}{eq:int_c}. 
Here the common items, e.g., the oscillation probability, are omitted to simplify the description.

To estimate the number of IBD events $N_i$ in a particular \Erec energy bin $i$, each group uses a different method of integration.
Group A computes the triple integral \eqref{eq:int_a}
of the differential IBD cross section $d\sigma/d\cos\theta$ and reactor antineutrino spectrum $S$ over the neutrino
energy \Enu, reconstructed energy \Erec, and $\cos\theta$. 
It considers the energy smearing due to neutron recoil via the scattering angle
of the positron $\cos\theta$, detector energy resolution, and LSNL within conversion to reconstructed energy $\Erec(\Evis(\Edep(\Epos(\Enu, \cos\theta))))$.
Group B uses the double integral \eqref{eq:int_b}, which considers the neutron recoil-related energy smearing and LSNL, defined within the conversion of neutrino energy to visible energy $\Evis(\Edep(\Epos(\Enu, \cos\theta)))$. 
The energy resolution is considered via the energy smearing matrix $\Ceres$. 
Group C also computes the double integral \eqref{eq:int_c}, which considers the energy conversion from neutrino energy to deposited energy. 
The LSNL and energy resolution are applied via the detector response matrix~$C$. 

{
  \renewcommand{\arraystretch}{3}%
  \newcommand\EqManual[1]{\refstepcounter{equation}\label{#1}(\theequation)}%
  \centering%
  \scalebox{0.95}{
    \begin{tabular}{
        @{}l                                  
        >{$\displaystyle}r<{$}@{}             
        >{$\displaystyle}l<{$}@{}             
        >{$\displaystyle}l<{$}
        r                                     
      }
      Group A:                     & N_i=                     &                                  & \int\limits_{\Erec_{i}}^{\Erec_{i+1}} d\Erec \int\limits_{1.8\text{ MeV}}^{15\text{ MeV}} d\Enu  \int\limits_{-1}^{1}d\cos\theta\  S(\Enu) \frac{d\sigma}{d\cos\theta}(\Enu,\cos\theta)  R(\Erec|\Enu),                             & \EqManual{eq:int_a}           \\
      Group B:                     & N_i=                     & \sum\limits_j \Ceres_{ij}        & \int\limits_{\Evis_{j}}^{\Evis_{j+1}} d\Evis                                                     \int\limits_{-1}^{1}d\cos\theta\  S(\Enu) \frac{d\sigma}{d\cos\theta}(\Enu,\cos\theta)  \frac{d\Enu}{d\Edep} \frac{d\Edep}{d\Evis}, & \EqManual{eq:int_b}           \\
      Group C:                     & N_i=                     & \sum\limits_j C_{ij}             & \int\limits_{\Edep_j}^{\Edep_{j+1}} d\Edep \int\limits_{1.8~\text{MeV}}^{15~\text{MeV}} d\Enu S(\Enu) \frac{d\sigma(\Enu, \Edep)}{d\Edep}.                                                                  & \EqManual{eq:int_c}           \\
    \end{tabular}
  }
}

The matrices $\Ceres$ and $C$ are built analytically based on the Gaussian smearing with the resolution~\eqref{eq:EnergyResolution:abc} and interpolated LSNL curves (see~\cref{subsec:lsnl}).
The IBD cross section, reactor antineutrino spectrum, and relevant systematic uncertainties are shared between the JUNO and TAO detectors.
The integration is performed within bins with a size of 5~keV, and the detector effects are applied as matrices defined on the same intervals. 
Finally, the histograms are rebinned into the final 340 intervals to estimate the number of IBD events in each energy bin.

For the statistical methods, Group A is using \cref{eq:chi2-a-b}, as discussed in \cref{subsec:stat}.
Group B uses the same formula, but with the regular Pearson's definition of the statistical uncertainties $\left(V\tstat\right)_{ii} = \Theory_i$ instead of CNP. For the nuisance term for the oscillation parameter, \SinSqDTheta13 is used instead of \SinSqTheta13.

Group C uses the ratio of Poisson functions for the statistical part of the test statistic:
\begin{multline}
  \TestStat{}_C(\ParsFree, \SinSqDTheta13, \ParsNuisance, \ParsNuisanceCorr)=
  2
  \sum_{\IdxDet,i}
  \left(\Theory\TDet_i(\ParsFree, \SinSqDTheta13, \ParsNuisance, \ParsNuisanceCorr)-\Data{}\TDet_i+\Data{}\TDet_i\log \frac{\Data{}\TDet_i}{\Theory\TDet}\right)
  + \\
  + \ChiSqOsc(\SinSqDTheta13)
  + \ChiSqNuisance(\ParsNuisance)
  + \ChiSqNuisanceCorr(\ParsNuisanceCorr).
\end{multline}
The nuisance part in this case also contains all bin-to-bin uncertainties.

\subfileref

\cleardoublepage
\bibliographystyle{h-physrev5}
\bibliography{references}
\end{document}